\documentclass{emulateapj}

\usepackage{lineno}
\usepackage{graphicx, subfigure}
\usepackage{amssymb}
\usepackage{amsmath}
\usepackage{epstopdf}
 \usepackage{hyperref}
\usepackage{color}
\usepackage{placeins}
\usepackage{natbib}
\usepackage{enumerate}
\usepackage{textcomp}
\usepackage{footmisc}
\usepackage{xspace}

\newcommand{\Fermi}{\emph{Fermi}\xspace}

\newcommand{\hsi}{\emph{RHESSI}\xspace}
\newcommand{\goes}{\emph{GOES}\xspace}

\newcommand{\stereo}{\emph{STEREO}\xspace}
\newcommand{\sdo}{\emph{SDO}\xspace}
\newcommand{\konus}{Konus-\emph{Wind}\xspace}

\newcommand{\PGSTAT}{\texttt{PGSTAT}\xspace}

\def\Rs{R$_\odot$\xspace}
\def\p0{$\pi^{\rm 0}$}
\def\Angst{$\buildrel _{\circ} \over {\mathrm{A}}$}
\def\de{$^{\circ}$\xspace}

\def\d{\delta}

\def\p8{\texttt{Pass8}}
\def\p7{\texttt{Pass7REP}}

\def\coul{{\rm Coul}}

\def\tsc{\tau_{\rm sc}\,}

\def\tcross{\tau_{\rm cross}\,}

\shorttitle{Gamma-rays from a Behind-the-limb Solar Flare}
\shortauthors{\Fermi-LAT Collaboration }

\begin{document}

\title{\Fermi-LAT observations of high-energy behind-the-limb solar flares}

\author{
M.~Ackermann\altaffilmark{1}, 
A.~Allafort\altaffilmark{2}, 
L.~Baldini\altaffilmark{3}, 
G.~Barbiellini\altaffilmark{4,5}, 
D.~Bastieri\altaffilmark{6,7}, 
R.~Bellazzini\altaffilmark{8}, 
E.~Bissaldi\altaffilmark{9}, 
R.~Bonino\altaffilmark{10,11}, 
E.~Bottacini\altaffilmark{2}, 
J.~Bregeon\altaffilmark{12}, 
P.~Bruel\altaffilmark{13}, 
R.~Buehler\altaffilmark{1}, 
R.~A.~Cameron\altaffilmark{2}, 
M.~Caragiulo\altaffilmark{14,9}, 
P.~A.~Caraveo\altaffilmark{15}, 
E.~Cavazzuti\altaffilmark{16}, 
C.~Cecchi\altaffilmark{17,18}, 
E.~Charles\altaffilmark{2}, 
S.~Ciprini\altaffilmark{16,17}, 
F.~Costanza\altaffilmark{9}, 
S.~Cutini\altaffilmark{16,17}, 
F.~D'Ammando\altaffilmark{19,20}, 
F.~de~Palma\altaffilmark{9,21}, 
R.~Desiante\altaffilmark{10,22}, 
S.~W.~Digel\altaffilmark{2}, 
N.~Di~Lalla\altaffilmark{3}, 
M.~Di~Mauro\altaffilmark{2}, 
L.~Di~Venere\altaffilmark{14,9}, 
P.~S.~Drell\altaffilmark{2}, 
C.~Favuzzi\altaffilmark{14,9}, 
Y.~Fukazawa\altaffilmark{23}, 
P.~Fusco\altaffilmark{14,9}, 
F.~Gargano\altaffilmark{9}, 
N.~Giglietto\altaffilmark{14,9}, 
F.~Giordano\altaffilmark{14,9}, 
M.~Giroletti\altaffilmark{19}, 
I.~A.~Grenier\altaffilmark{24}, 
L.~Guillemot\altaffilmark{25,26}, 
S.~Guiriec\altaffilmark{27,28}, 
T.~Jogler\altaffilmark{29}, 
G.~J\'ohannesson\altaffilmark{30,54}, 
L.~Kashapova\altaffilmark{31}, 
S.~Krucker\altaffilmark{32,33}, 
M.~Kuss\altaffilmark{8}, 
G.~La~Mura\altaffilmark{7}, 
S.~Larsson\altaffilmark{34,35}, 
L.~Latronico\altaffilmark{10}, 
J.~Li\altaffilmark{36}, 
W.~Liu\altaffilmark{37,38,2}, 
F.~Longo\altaffilmark{4,5}, 
F.~Loparco\altaffilmark{14,9}, 
P.~Lubrano\altaffilmark{17}, 
J.~D.~Magill\altaffilmark{39}, 
S.~Maldera\altaffilmark{10}, 
A.~Manfreda\altaffilmark{3}, 
M.~N.~Mazziotta\altaffilmark{9}, 
W.~Mitthumsiri\altaffilmark{40}, 
T.~Mizuno\altaffilmark{41}, 
M.~E.~Monzani\altaffilmark{2}, 
A.~Morselli\altaffilmark{42}, 
I.~V.~Moskalenko\altaffilmark{2}, 
M.~Negro\altaffilmark{10,11}, 
E.~Nuss\altaffilmark{12}, 
T.~Ohsugi\altaffilmark{41}, 
N.~Omodei\altaffilmark{2,43}, 
E.~Orlando\altaffilmark{2}, 
V.~Pal'shin\altaffilmark{44}, 
D.~Paneque\altaffilmark{45}, 
J.~S.~Perkins\altaffilmark{27}, 
M.~Pesce-Rollins\altaffilmark{8,46}, 
V.~Petrosian\altaffilmark{2,47}, 
F.~Piron\altaffilmark{12}, 
G.~Principe\altaffilmark{48}, 
S.~Rain\`o\altaffilmark{14,9}, 
R.~Rando\altaffilmark{6,7}, 
M.~Razzano\altaffilmark{8,49}, 
O.~Reimer\altaffilmark{50,2}, 
F.~Rubio~da~Costa\altaffilmark{2}, 
C.~Sgr\`o\altaffilmark{8}, 
D.~Simone\altaffilmark{9}, 
E.~J.~Siskind\altaffilmark{51}, 
F.~Spada\altaffilmark{8}, 
G.~Spandre\altaffilmark{8}, 
P.~Spinelli\altaffilmark{14,9}, 
H.~Tajima\altaffilmark{52,2}, 
J.~B.~Thayer\altaffilmark{2}, 
D.~F.~Torres\altaffilmark{36,53}, 
E.~Troja\altaffilmark{27,39}, 
G.~Vianello\altaffilmark{2}
}
\altaffiltext{1}{Deutsches Elektronen Synchrotron DESY, D-15738 Zeuthen, Germany}
\altaffiltext{2}{W. W. Hansen Experimental Physics Laboratory, Kavli Institute for Particle Astrophysics and Cosmology, Department of Physics and SLAC National Accelerator Laboratory, Stanford University, Stanford, CA 94305, USA}
\altaffiltext{3}{Universit\`a di Pisa and Istituto Nazionale di Fisica Nucleare, Sezione di Pisa I-56127 Pisa, Italy}
\altaffiltext{4}{Istituto Nazionale di Fisica Nucleare, Sezione di Trieste, I-34127 Trieste, Italy}
\altaffiltext{5}{Dipartimento di Fisica, Universit\`a di Trieste, I-34127 Trieste, Italy}
\altaffiltext{6}{Istituto Nazionale di Fisica Nucleare, Sezione di Padova, I-35131 Padova, Italy}
\altaffiltext{7}{Dipartimento di Fisica e Astronomia ``G. Galilei'', Universit\`a di Padova, I-35131 Padova, Italy}
\altaffiltext{8}{Istituto Nazionale di Fisica Nucleare, Sezione di Pisa, I-56127 Pisa, Italy}
\altaffiltext{9}{Istituto Nazionale di Fisica Nucleare, Sezione di Bari, I-70126 Bari, Italy}
\altaffiltext{10}{Istituto Nazionale di Fisica Nucleare, Sezione di Torino, I-10125 Torino, Italy}
\altaffiltext{11}{Dipartimento di Fisica, Universit\`a degli Studi di Torino, I-10125 Torino, Italy}
\altaffiltext{12}{Laboratoire Univers et Particules de Montpellier, Universit\'e Montpellier, CNRS/IN2P3, F-34095 Montpellier, France}
\altaffiltext{13}{Laboratoire Leprince-Ringuet, \'Ecole polytechnique, CNRS/IN2P3, F-91128 Palaiseau, France}
\altaffiltext{14}{Dipartimento di Fisica ``M. Merlin" dell'Universit\`a e del Politecnico di Bari, I-70126 Bari, Italy}
\altaffiltext{15}{INAF-Istituto di Astrofisica Spaziale e Fisica Cosmica Milano, via E. Bassini 15, I-20133 Milano, Italy}
\altaffiltext{16}{Agenzia Spaziale Italiana (ASI) Science Data Center, I-00133 Roma, Italy}
\altaffiltext{17}{Istituto Nazionale di Fisica Nucleare, Sezione di Perugia, I-06123 Perugia, Italy}
\altaffiltext{18}{Dipartimento di Fisica, Universit\`a degli Studi di Perugia, I-06123 Perugia, Italy}
\altaffiltext{19}{INAF Istituto di Radioastronomia, I-40129 Bologna, Italy}
\altaffiltext{20}{Dipartimento di Astronomia, Universit\`a di Bologna, I-40127 Bologna, Italy}
\altaffiltext{21}{Universit\`a Telematica Pegaso, Piazza Trieste e Trento, 48, I-80132 Napoli, Italy}
\altaffiltext{22}{Universit\`a di Udine, I-33100 Udine, Italy}
\altaffiltext{23}{Department of Physical Sciences, Hiroshima University, Higashi-Hiroshima, Hiroshima 739-8526, Japan}
\altaffiltext{24}{Laboratoire AIM, CEA-IRFU/CNRS/Universit\'e Paris Diderot, Service d'Astrophysique, CEA Saclay, F-91191 Gif sur Yvette, France}
\altaffiltext{25}{Laboratoire de Physique et Chimie de l'Environnement et de l'Espace -- Universit\'e d'Orl\'eans / CNRS, F-45071 Orl\'eans Cedex 02, France}
\altaffiltext{26}{Station de radioastronomie de Nan\c{c}ay, Observatoire de Paris, CNRS/INSU, F-18330 Nan\c{c}ay, France}
\altaffiltext{27}{NASA Goddard Space Flight Center, Greenbelt, MD 20771, USA}
\altaffiltext{28}{NASA Postdoctoral Program Fellow, USA}
\altaffiltext{29}{Friedrich-Alexander-Universit\"at, Erlangen-N\"urnberg, Schlossplatz 4, 91054 Erlangen, Germany}
\altaffiltext{30}{Science Institute, University of Iceland, IS-107 Reykjavik, Iceland}
\altaffiltext{31}{Institute of Solar-Terrestrial SB RAS, Lermontov st. 126a, 6640333, Irkutsk, Russia}
\altaffiltext{32}{University of Applied Sciences and Arts Northwestern Switzerland, CH-5210 Windisch, Switzerland}
\altaffiltext{33}{Space Science Laboratory, University of California, Berkeley, CA 94720-7450, USA}
\altaffiltext{34}{Department of Physics, KTH Royal Institute of Technology, AlbaNova, SE-106 91 Stockholm, Sweden}
\altaffiltext{35}{The Oskar Klein Centre for Cosmoparticle Physics, AlbaNova, SE-106 91 Stockholm, Sweden}
\altaffiltext{36}{Institute of Space Sciences (IEEC-CSIC), Campus UAB, Carrer de Magrans s/n, E-08193 Barcelona, Spain}
\altaffiltext{37}{Bay Area Environmental Research Institute, 625 2nd Street, Suite 209, Petaluma, CA 94952, USA}
\altaffiltext{38}{Lockheed Martin Solar and Astrophysics Laboratory, 3251 Hanover Street, Bldg. 252, Palo Alto, CA 94304, USA}
\altaffiltext{39}{Department of Physics and Department of Astronomy, University of Maryland, College Park, MD 20742, USA}
\altaffiltext{40}{Department of Physics, Faculty of Science, Mahidol University, Bangkok 10400, Thailand}
\altaffiltext{41}{Hiroshima Astrophysical Science Center, Hiroshima University, Higashi-Hiroshima, Hiroshima 739-8526, Japan}
\altaffiltext{42}{Istituto Nazionale di Fisica Nucleare, Sezione di Roma ``Tor Vergata", I-00133 Roma, Italy}
\altaffiltext{43}{email: nicola.omodei@stanford.edu}
\altaffiltext{44}{St. Peterburg, Vedeneeva 2-31, Russia}
\altaffiltext{45}{Max-Planck-Institut f\"ur Physik, D-80805 M\"unchen, Germany}
\altaffiltext{46}{email: melissa.pesce.rollins@pi.infn.it}
\altaffiltext{47}{email: vahep@stanford.edu}
\altaffiltext{48}{Erlangen Centre for Astroparticle Physics, D-91058 Erlangen, Germany}
\altaffiltext{49}{Funded by contract FIRB-2012-RBFR12PM1F from the Italian Ministry of Education, University and Research (MIUR)}
\altaffiltext{50}{Institut f\"ur Astro- und Teilchenphysik and Institut f\"ur Theoretische Physik, Leopold-Franzens-Universit\"at Innsbruck, A-6020 Innsbruck, Austria}
\altaffiltext{51}{NYCB Real-Time Computing Inc., Lattingtown, NY 11560-1025, USA}
\altaffiltext{52}{Solar-Terrestrial Environment Laboratory, Nagoya University, Nagoya 464-8601, Japan}
\altaffiltext{53}{Instituci\'o Catalana de Recerca i Estudis Avan\c{c}ats (ICREA), E-08010 Barcelona, Spain}
\altaffiltext{54}{NORDITA,  Roslagstullsbacken 23, 106 91 Stockholm, Sweden}

\begin{abstract}
We report on the \Fermi-LAT detection of high-energy emission from the
behind-the-limb solar flares that occurred on 2013 October 11, 2014 January 6 and 2014 September 1. The \Fermi-LAT observations are associated with flares from active regions originating behind both the eastern and western limbs, as determined by \stereo. All three flares are associated with very fast Coronal Mass Ejections (CMEs) and strong Solar Energetic Particle events. We present
updated localizations of the $>100$ MeV photon emission, hard X-ray (HXR) and EUV images and broad-band spectra from 10 keV to 10 GeV as well as microwave spectra.  We also
provide a comparison of the behind-the-limb flares detected by the \Fermi-LAT
with three on-disk flares and present a study of some of the significant quantities
of these flares as an attempt to better understand the acceleration mechanisms at
work during these occulted flares. We interpret the HXR emission to be due to
electron bremsstrahlung from a {\it coronal} thin-target loop-top with the accelerated
electron spectra steepening at semirelativistic energies. The $>100$
MeV gamma-rays are best described by a pion decay model resulting from
interaction of protons (and other ions) in a thick target photospheric source.
The protons are believed to have been accelerated  (to energies $> 10$ GeV) in the CME environment and precipitate down to the photosphere from the down-stream side of the CME shock and landed on the front side of the Sun, away from the original flare site and the HXR emission.
\end{abstract}

\keywords{Sun: flares: Sun: X-rays, gamma rays}

\section{Introduction}

Gamma-ray emission from solar flares is generally believed to occur
predominantly in compact high-density regions near the photospheric footpoints of
magnetic field lines.
Observations of gamma-ray emission from flares whose host active regions are located
behind the visible solar disk pose interesting questions regarding the acceleration sites and mechanism,  the transport and interaction points of the accelerated particles during these rare events.

Three behind-the-limb (BTL)  flares with emission up to 100~MeV were observed during solar cycles 21 and 22. The first occulted solar flare whose
active region (AR) was estimated to be 15$^{\circ}$ behind the western limb was
observed on 1989 September 29 by the Gamma-Ray Spectrometer (GRS) on-board the {\it Solar Maximum Mission (SMM)}.
\citet{1993ApJ...409L..69V} reported intense gamma-ray line emission in the 1--8
MeV range and a strong 2.223 MeV neutron capture line from this flare. Given the strength of the line emission it was concluded that a spatially extended component was required in order to explain the observations. The second, detected by PHEBUS on {\it GRANAT}~\citep{1994ApJ...425L.109B} on 1991 June 1, had intense
gamma-ray line emission in the 1--8 MeV range but no  neutron capture line,
indicating that the emission was of coronal origin. The third occulted flare, detected by PHEBUS, BATSE and EGRET on 1991 June 30, was an electron-dominated flare with no detectable line emission but with significant emission up to almost 100 MeV.
\citet{1999A&A...342..575V} report that the spectral properties of this flare
were similar to those of  the flares occurring  on the visible disk. Although
there was some speculation and some scenarios were put forth~\citep{cliv93}, no definite explanations were found on  how and where the particles responsible for these emissions were accelerated and where the gamma-rays were produced.

Hard X-ray (HXR) emissions from the loop-top of flares originating from ARs located just beyond the limb is often observed by \hsi~\citep{sald08}. These are  referred to as partially occulted flares since only the emission from the loop footpoints are occulted. An unusual such flare was reported by~\citet{KruckerS_Coronal60keV_2007ApJ...669L..49K} where the flare originated 40$^{\circ}$ behind the limb.  This is very  similar to the 2014 September 1 flare discussed in this paper.

The \Fermi-Large Area Telescope (LAT)~\citep{LATPaper} observations have doubled the number of occulted flares detected and provided the first detections of emission in the GeV range from such rare events. The \Fermi-LAT observations sample flares from active regions originating from behind both the eastern and western limbs, are all associated with very fast Coronal Massive Ejections (CMEs) and strong Solar Energetic Particle (SEP) events.  In this paper we present the  observations of the first three BTL flares detected by \Fermi-LAT together with those from complementary instruments such as \stereo, \hsi, \konus~\citep{1995SSRv...71..265A}, \Fermi-GBM~\citep[GBM;][]{meeg09} as well as solar observatories in the Radio Solar Telescope Network (RSTN).

In \S~\ref{sec:btf}  we present the  multiwavelength features including the light curves and localization of the various emissions. Then in \S 3 we describe a more detailed analysis of the spectral
data and compare the properties of the BTL flares with three on-disk flares
observed by the \Fermi-LAT. In \S 4 we discuss the associated SEP observations
and in \S 5 we present a brief summary and interpretation of these results.

\section{Observations of \Fermi-LAT behind-the-limb flares}
\label{sec:btf}

We have collected all relevant observations of the three BTL flares
detected by the \Fermi-LAT. In what follows we present these observations. A
first analysis of the SOL2013-10-11 flare based on Pass7\_REP data was presented in~\citet{2015ApJ...805L..15P}. We have re-analyzed  all three of the flares using Pass 8 data. The Pass 8 data benefit\footnote{A summary of the \Fermi-LAT Pass 8 performance can be found at \url{http://www.slac.stanford.edu/exp/glast/groups/canda/lat\_Performance.htm}} from an improved point-spread function, effective area, and energy reach.

\subsection{Light Curves}
\label{sec:LC}
\begin{figure}[htb!]
\begin{center}
\includegraphics[trim=1cm 3cm 2cm 2cm, clip=true, width=0.48\textwidth]{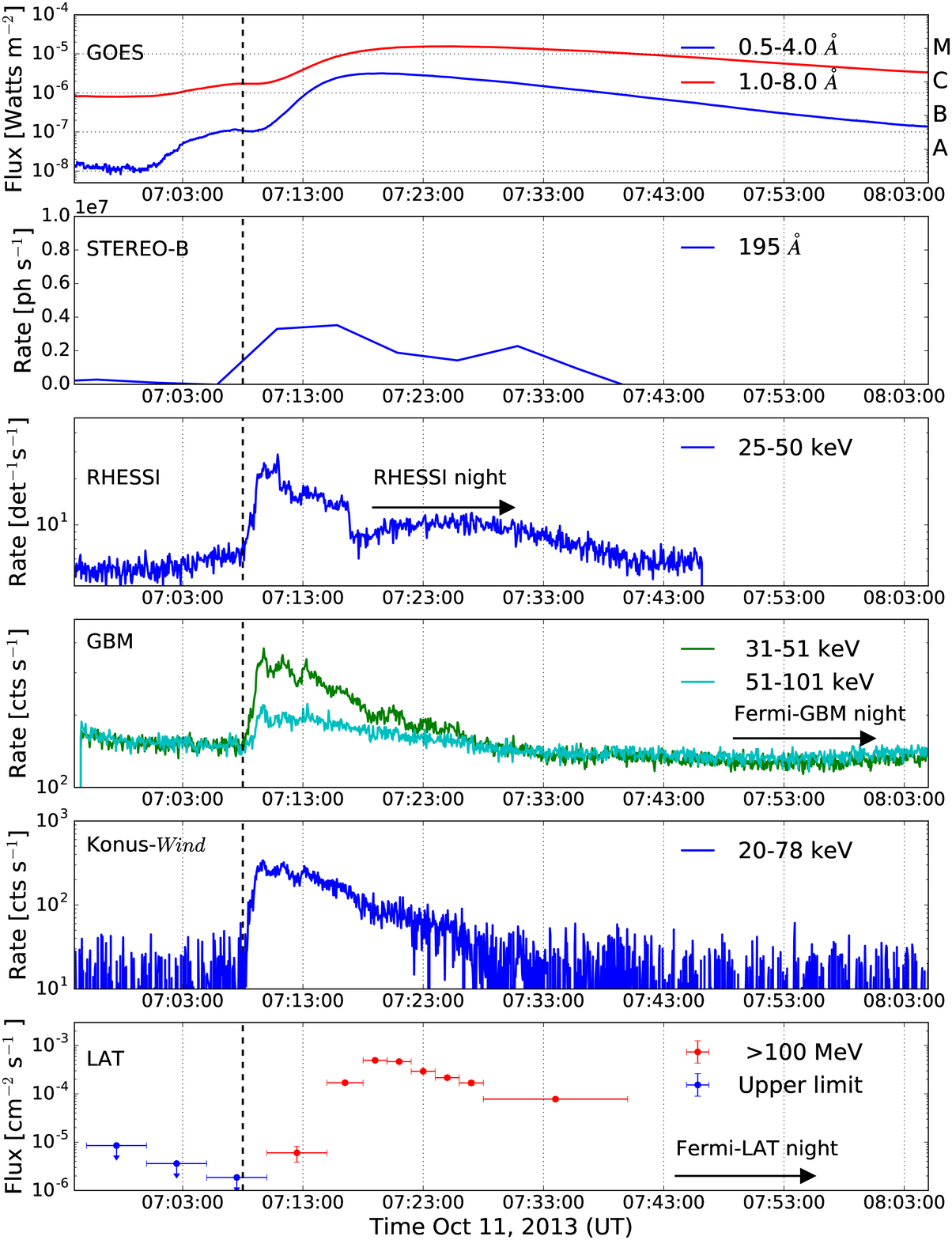}
\caption{Composite light curve for Oct13 with data from \goes, \stereo, \hsi, \Fermi-GBM, \konus and \Fermi-LAT. The vertical dashed line represents the estimated start time of the flare, 7:08 UT based on \stereo-B observations. \hsi night started at around 07:15 UT and \Fermi-LAT night started around 07:43 UT.}
\label{fig:SOL20131011_LC}
\end{center}
\end{figure}

{\it SOL2013-10-11:} On 2013 October 11 (Oct13) between 07:05:51 and 07:10:51 UT \stereo-B detected a solar flare with an AR located at N21E103. \goes detected an M1.5 class flare starting at around the same time as \stereo-B. However, based on the \stereo-B 195 \Angst\ emission and the method described in~\citet{2013SoPh..288..241N} we estimate that the \goes class of this flare if the active region had not been occulted would have been M4.9 with an uncertainty within a factor of three. This method utilizes the pre-flare background subtracted, full-disk integrated EUV intensity, as shown in Figure~\ref{fig:SOL20131011_LC}b. A fast CME was observed by LASCO with a reported first appearance by the white light coronagraph C2 (imaging from 2-6 solar radii) at 07:24:10 UT and a linear speed of 1200 km s$^{-1}$.
\Fermi coverage started at 07:08:00 UT and continued for more than 30 minutes. The \Fermi-LAT detected $>$100~MeV emission for $\sim$30 min with the maximum of the flux occurring between 07:20:00--07:25:00 UT\footnote{From the re-analysis of this flare  with Pass 8 data we gain 5 minutes of additional emission with respect to our previous work~\citep{2015ApJ...805L..15P}.}. The \Fermi-GBM detection of HXRs began a few minutes before the \Fermi-LAT and peaked earlier ($\sim$07:10 above 50 keV). \hsi coverage was from 07:08:00--07:16:40 UT, overlapping with \Fermi for 9 min.  \konus, located at Lagrangian point L1, detected emission in the 20-78~keV and 78-310~keV energy bands simultaneously with \hsi and \Fermi-GBM. This  flare was also observed in radio by the Radio Solar Telescope Network (RSTN) and the Nobeyama radio polarimeter~\citep{1994IEEEP..82..705N} at frequencies up to 9 GHz (see Section~\ref{sec:radio}). 

\begin{figure}[htb!]
\begin{center}
  \includegraphics[width=0.46\textwidth]{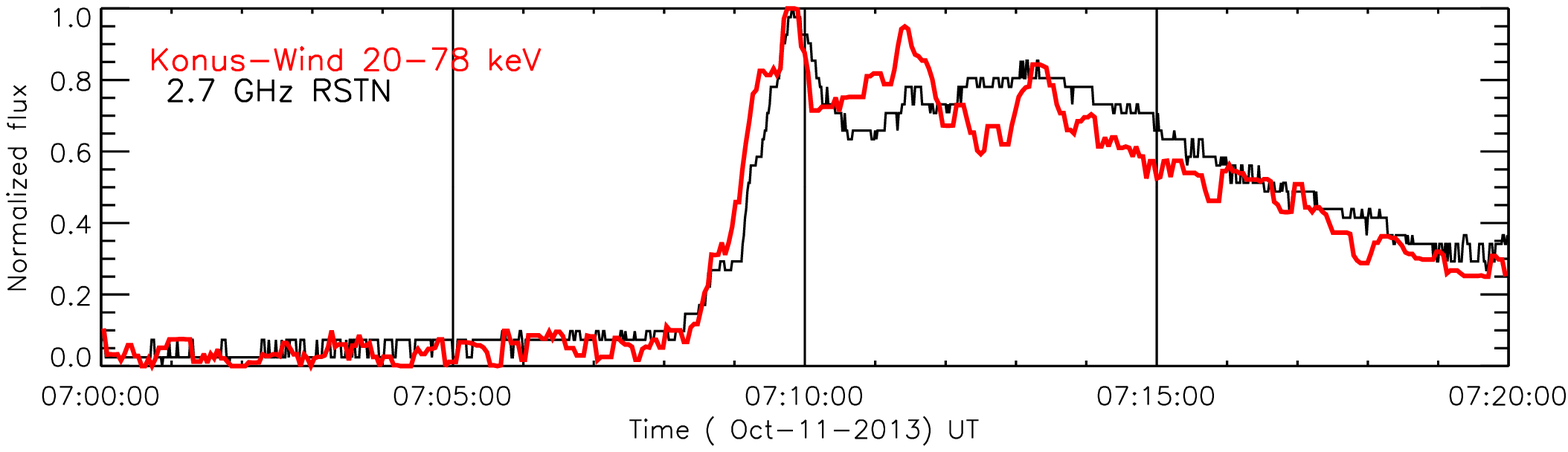}\\
  \includegraphics[width=0.46\textwidth]{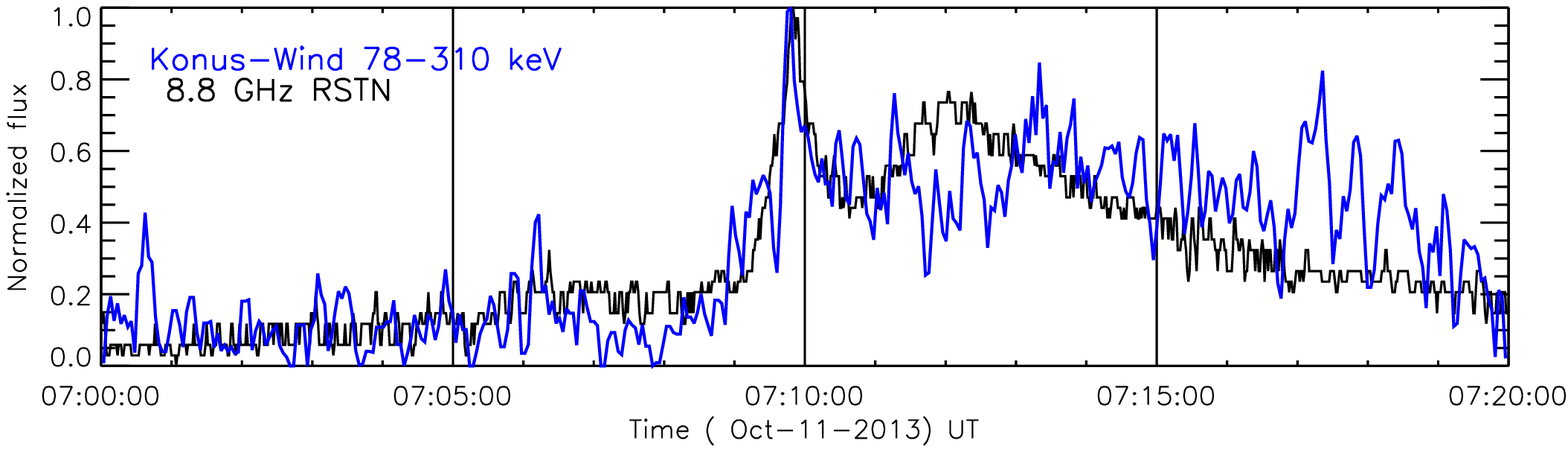}
\caption{Comparison of the time profiles of HXR and microwave (MW) emission for Oct13. Top and bottom panels compare the 20-78 keV HXR light curve with optically thick gyro-synchrotron (GS) emission at 2.7 GHz and 78-310 keV HXR light curve with the optically thin GS emission at 8.8 GHz, respectively. Note that the shape of the GS spectrum it is such that for frequencies below the peak the spectrum is considered optically thick whereas above the peak is it optically thin.}  
\label{fig:KonusAndRadioSOL2013-10-11}
\end{center}
\end{figure}

The light curves (LCs) for the \goes, \stereo-B, \hsi, \Fermi-GBM,  \konus and \Fermi-LAT are shown in Figure~\ref{fig:SOL20131011_LC}. The microwave (MW) emission is compared with the HXR light curve from \konus in Figure~\ref{fig:KonusAndRadioSOL2013-10-11}.

{\it SOL2014-01-06:} On 2014 January 6 (Jan14) between 07:35:46 and 07:45:46 UT a solar flare erupted from an AR located at S8W110. Both \stereo spacecrafts had a full view of the AR and detected a large filament eruption from the AR at approximately 07:50:00 UT. The tip of this filament was seen from the visible solar disk by \sdo/AIA. There is a hint of detection by \emph{GOES} of a smaller-than-C-class flare. However, the peak rate of 2.5$\times$10$^{5}$ photons s$^{-1}$ detected by \stereo-B in its 195 \Angst\ channel indicates that the flare would have been classified as  a \emph{GOES} X3.5 if it had not been occulted~\citep{2013SoPh..288..241N}. LASCO detected a halo CME with a first C2 appearance at 08:00:05 UT with a linear speed of 1400 km s$^{-1}$.

\begin{figure}[htb!]
\begin{center}
\includegraphics[trim=1cm 3cm 2cm 2cm, clip=true, width=0.48\textwidth]{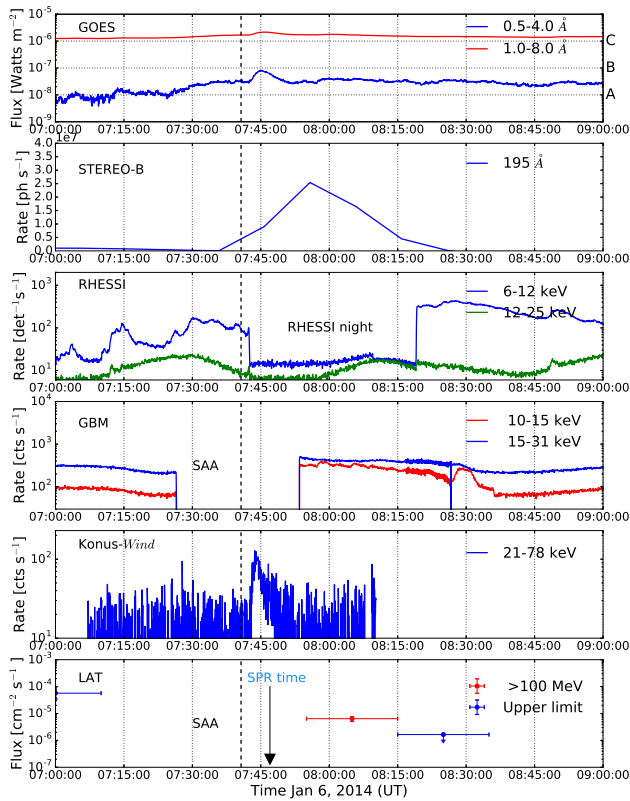}
\caption{Composite light curve for Jan14 with data from GOES, \stereo, \hsi, \Fermi-GBM, \konus and \Fermi-LAT. The vertical dashed line represents the estimated start time of the flare, 07:40:50 UT based on \stereo-B emission. The time of Solar Particle Release (SPR) in the solar atmosphere is described in Section~\ref{sec:Solar Energetic Particles}.}
\label{fig:SOL20140106_LC}
\end{center}
\end{figure}

Upon exiting the South Atlantic Anomaly (SAA)\footnote{Both detectors on-board \Fermi\ are turned off while the spacecraft is in the SAA.} at 07:55:00 UT both instruments on-board \Fermi detected emission associated with this flare. The \Fermi-LAT detected $>$100 MeV emission for approximately 20 minutes (with no evidence of temporally extended emission after 08:15:00 UT.), and the \Fermi-GBM detected emission in the tens of keV range.  \hsi detected emission starting at $\sim$08:18:00 UT (upon exiting spacecraft night) also in the tens of keV energy range for over 40 minutes. \konus had a full view of the flare and detected emission only in the softest energy band, 20--78 keV, starting at 07:43:00 UT. Radio data above 1.3 GHz from RSTN for this flare did not indicate a detection. The  \goes, \hsi, \konus, \Fermi-GBM and \Fermi-LAT light curves for this flare are shown in Figure~\ref{fig:SOL20140106_LC}.

{\it SOL2014-09-01:} On 2014 September 1 (Sep14) between 10:55:56 and 11:00:56 UT a bright solar flare occurred in an AR located at N14E126. There was no \goes signal but  \stereo-B had an unblocked view of the entire flare and detected a maximum rate of 1.7$\times$10$^{7}$ photons s$^{-1}$ in its 195 \Angst\ channel, indicating an un-occulted \goes X2.4 class~\citep{2013SoPh..288..241N}. LASCO detected a halo CME with first C2 appearance at 11:12:05 UT with a linear speed of 1900 km s$^{-1}$. A Type II radio burst with an estimated velocity of 2079 km s$^{-1}$ was reported by NOAA Space Weather Alerts in association with this flare. \sdo/AIA reported a coronal wave from this AR starting during 10:45:35--12:21:35 UT. This wave was seen to propagate along the limb and over onto the visible disk\footnote{as reported by Nitta on \url{http://www.lmsal.com/isolsearch}}.
\begin{figure}[htb!]
\begin{center}
  \includegraphics[trim=1cm 2cm 2cm 2cm, clip=true, width=0.5\textwidth]{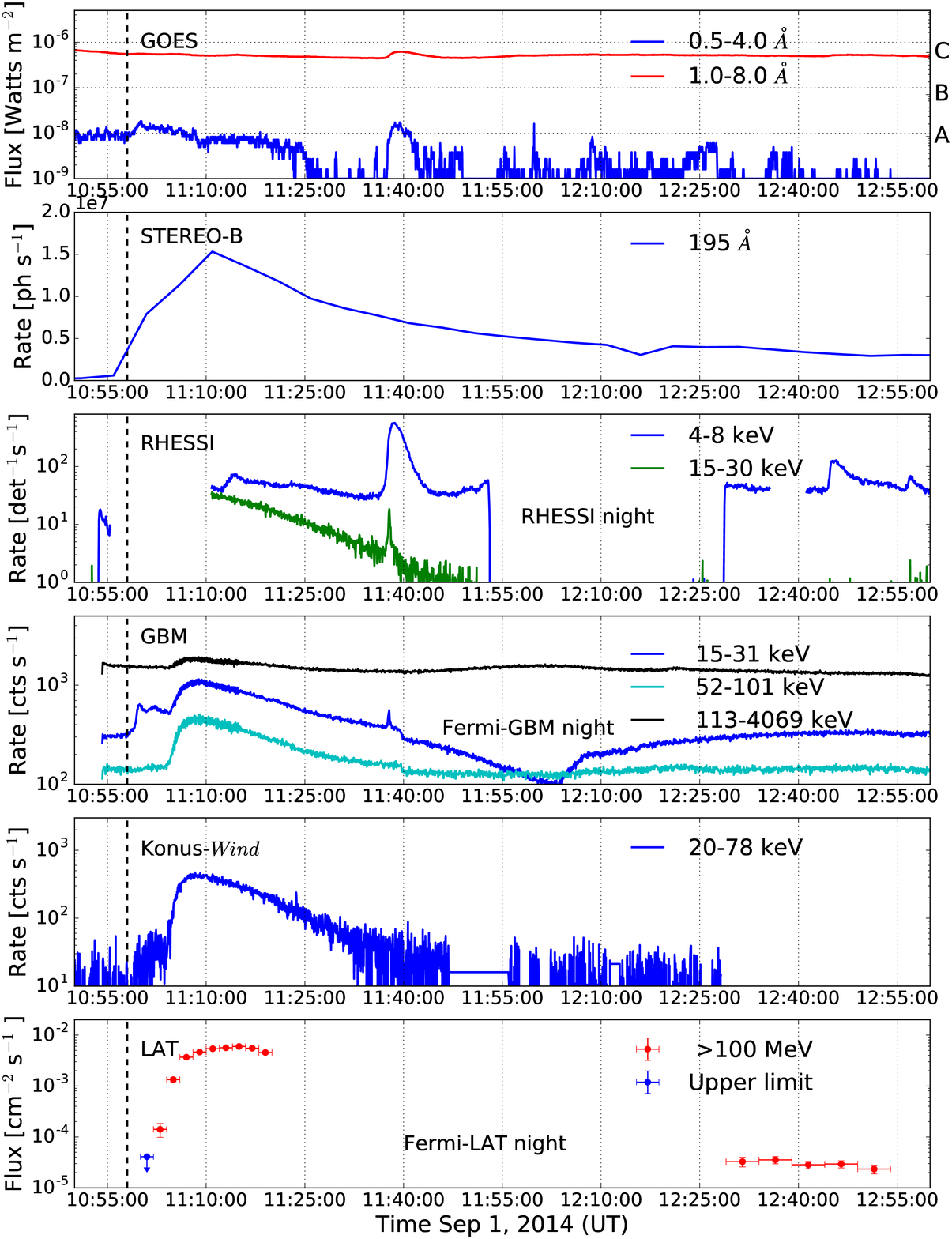}
\caption{Light curves of Sep14 as seen by \goes, \stereo, \hsi, \Fermi-GBM (black line is for the BGO detector while cyan and blue are for NaI), \konus and \Fermi-LAT. The vertical dashed line represents the estimated starting time of the flare, 10:58:00 UT based on \stereo-B emission. \Fermi-LAT night was from $\sim$11:35--12:20 UT. The increase in rate seen in \hsi and \Fermi-GBM at 11:38 UT is due to a small on-disk flare.}
\label{fig:SOL20140901LightCurve}
\end{center}
\end{figure}

\begin{figure}[htb!]
\begin{center}
  \includegraphics[width=0.46\textwidth]{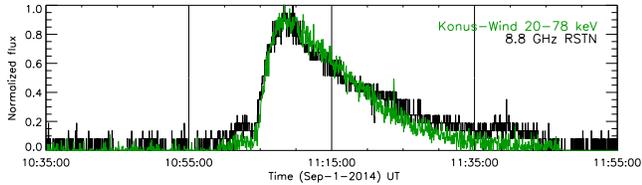}\\
\caption{Comparison of the 20--78 keV HXR emission and MW light curves at 8.8 GHz (optically thin gyro-synchrotron emission) for the Sep14 flare. Profiles are normalized to the peak value.}
\label{fig:KonusAndRadioSOL2014-09-01}
\end{center}
\end{figure}

The \Fermi-LAT detected emission from this flare for $\sim$2 hours, peaking between 11:10:00--11:15:00 UT\footnote{Thanks to improvements provided by the new \Fermi-LAT event selection we gained 10 extra minutes of coverage with respect to Pass7\_REP.}. The GBM detected emission up to a few MeV in temporal coincidence with the \Fermi-LAT emission in both the BGO and NaI detectors. \hsi was in the SAA from 10:55:00 to 11:11:00; upon exiting the SAA it detected emission up to 30 keV. \konus detected emission in all three energy bands: 20--78~keV, 78--310~keV and 310--1180~keV in temporal coincidence with \Fermi-GBM. A significant radio flux at frequencies up to 16 GHz was detected by the San Vito station of the RSTN simultaneously with the HXR emission peak detected by both \Fermi-GBM and \konus. Figure~\ref{fig:SOL20140901LightCurve} shows the light curves from GOES, \stereo, \hsi, \Fermi-GBM, \konus and \Fermi-LAT.  The light curves of the
microwave emission (MW) are compared with the HXR light curve from \konus
in Figure~\ref{fig:KonusAndRadioSOL2014-09-01}. The MW intensity distribution with frequency has a maximum  at about 1 GHz (see Section~\ref{sec:radio}). Thus we show the time profile of only this one frequency for this flare. As evident from Figure~\ref{fig:KonusAndRadioSOL2014-09-01} there is good agreement between these light curves.

\subsection{Localization of the Emission}
\label{sec:positions}

We present  composite images of the Sun as seen by \stereo and
\sdo and, whenever available, the position of HXR emission based on \hsi and
$>100$ MeV gamma-ray emission centroid based on \Fermi-LAT data.  We used the FITS World Coordinate System software package~\citep{2010SoPh..261..215T} to co-register the locations of the flares between \stereo\ and \sdo images.
We applied the CLEAN imaging algorithm~\citep{Hurford2002} to \hsi data using the detectors 3--9 to reconstruct the X-ray images. The centroid for the  $>$100 MeV gamma-ray emission is determined using the \texttt{gtfindsrc} tool, which performs a likelihood analysis of the average position in the time integrated data set.

\begin{figure*}[htb!]
\begin{center}
\includegraphics[width=0.8\textwidth]{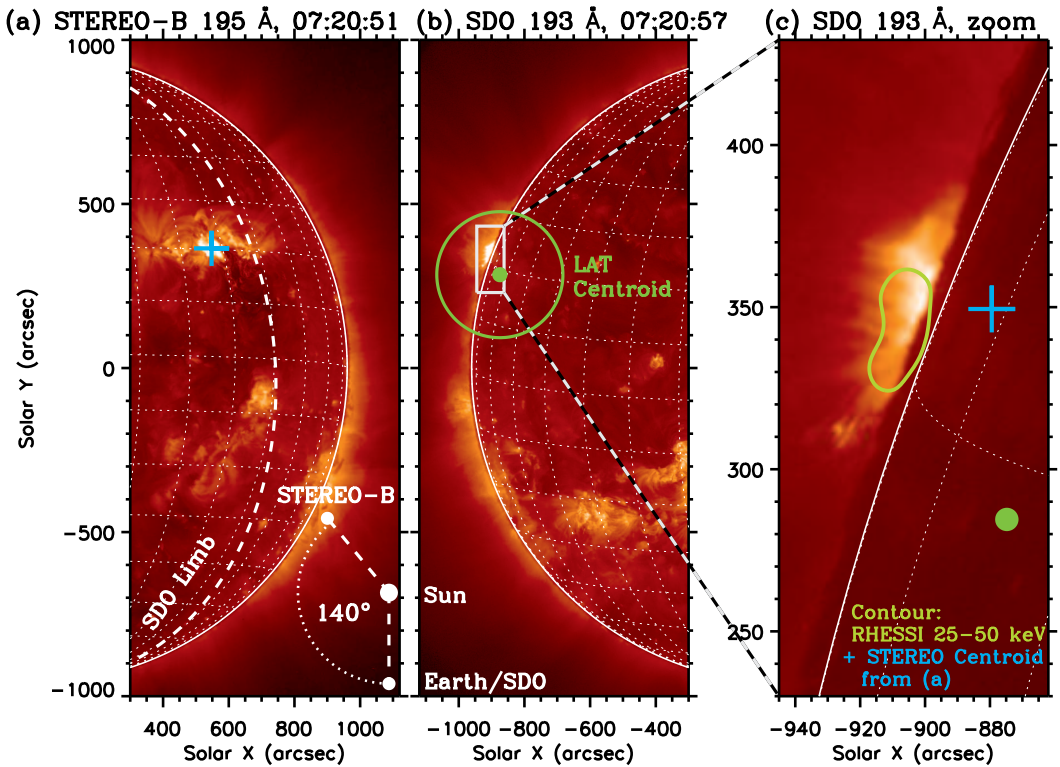}
  \caption{Localization of the Oct13 flare. Images near the flare peak are shown for \stereo-B 195 \Angst\ (a), \sdo 193  \Angst\ (b) and an enlargement of the \sdo image (c) marked by the white rectangle in (b).  The green circle in (b) shows the 68\% error
circle for the \Fermi-LAT emission centroid and the green dot in (c) represents the \Fermi-LAT emission centroid position. The time range for the \Fermi-LAT emission is from 07:10:00--07:35:00 UT. The green contour in (c) shows the 25--50 keV \hsi source. The blue cross in (a) and (c) mark the centroid of the \stereo\ flare ribbon as seen from the \stereo\ and Earth/\sdo\ perspectives, respectively. The white-dashed line in (a) represents the solar limb as seen by \sdo. The positions of \stereo\ and \sdo/Earth relative to the Sun are shown in the lower-right corner of (a).}
\label{fig:SOL20131011Loc}
\end{center}
\end{figure*}

Figure~\ref{fig:SOL20131011Loc} displays the composite image for the Oct13
flare showing the \stereo-B 195 \Angst\ image of the AR located about 10$^{\circ}$ behind the limb, the \sdo/AIA 193 \Angst\ emission peaking above the limb, a contour of the \hsi
image and results from  re-analysis of the flare with Pass 8 data. The
new emission centroid is located in
heliocentric coordinates [-880$\arcsec$, 290$\arcsec$], about 200$\arcsec$ closer to the \hsi centroid, and  with a 68\% error radius of
190$\arcsec$, which is $\sim$20\% smaller than the value we reported
in~\citet{2015ApJ...805L..15P} using Pass7\_REP data. In addition, in the Pass 8 data set, the total number of $>$1 GeV events measured from this flare increased from
4 to 7. The highest-energy photon detected from this flare was 3.4 GeV and the
arrival was 07:19:00 UT.

\begin{figure}[!hb]
\begin{center}
\includegraphics[width=0.48\textwidth]{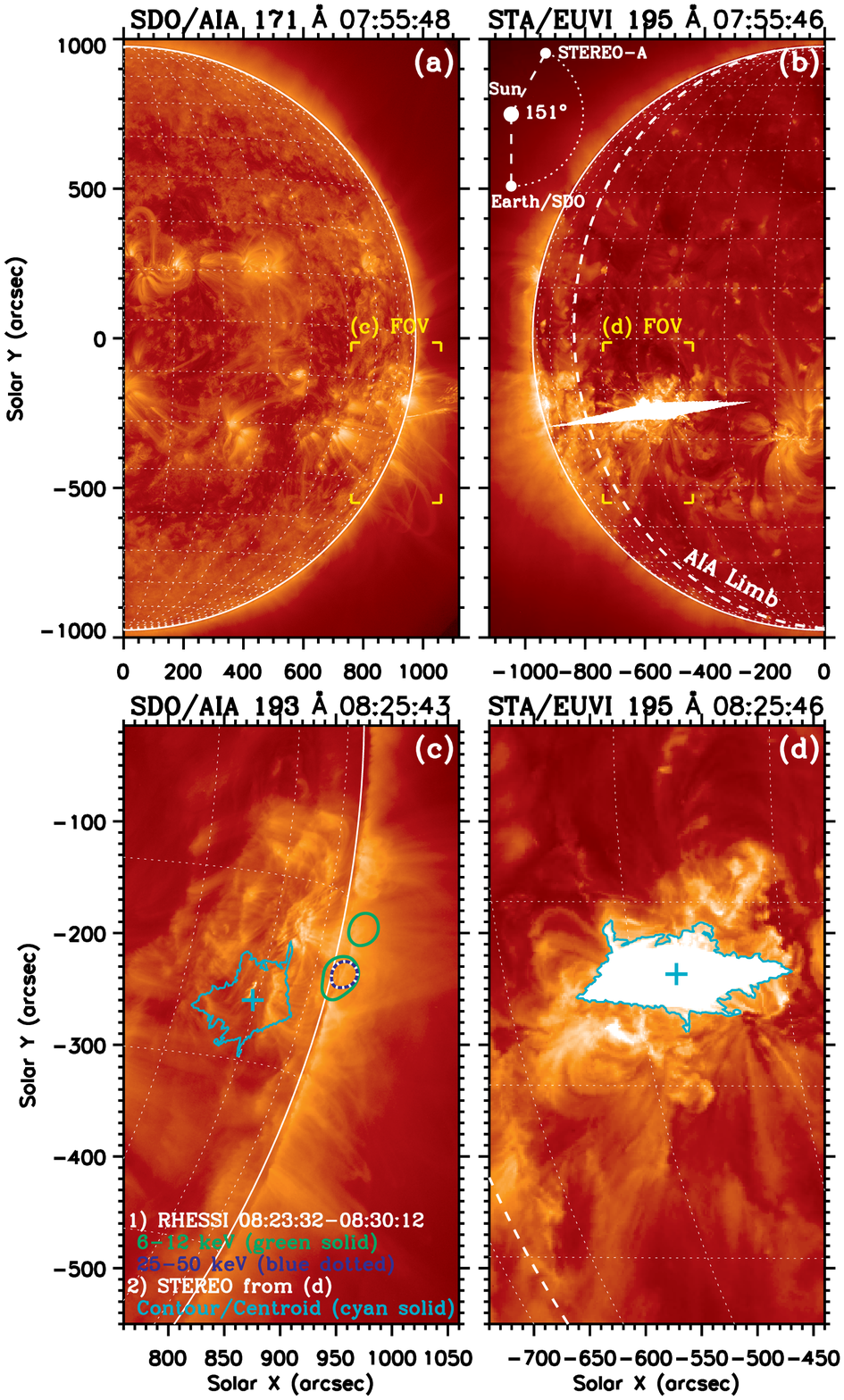}
\caption{\sdo 171 and 193 \Angst\ (left) and \stereo-B 195 \Angst\ (right) images near the Jan14 flare peak. The
white-dashed line in (b) and (d) represents the solar limb as seen by \sdo. The cyan contour and cross in (d) mark the \stereo\ flare
ribbon and its centroid, respectively. Their projected view as seen in AIA perspective is shown in (c), in which the centroid is located at 20\de behind the western limb. The green solid and blue dotted contours in (d) show the \hsi 6--12 and 25--50 keV sources, respectively. The rectangular brackets in (a) and (b) mark the field of view for (c) and (d), respectively.}
\label{fig:SOL20140106Loc}
\end{center}
\end{figure}

For the Jan14 flare the \Fermi-LAT photon statistics were not sufficient to provide an emission localization error circle smaller than 0.5$^{\circ}$. However, we can still conclude that the emission detected by the \Fermi-LAT was consistent with the position of the Sun.
In Figure~\ref{fig:SOL20140106Loc} we show the \stereo-A and \sdo images of this event at two different times. The top panels of Figure~\ref{fig:SOL20140106Loc} present \sdo 171 \Angst\ (left) and \stereo-A 195 \Angst\ (right) at 07:55:46 UT and show the filament eruption, while the bottom panels show the \sdo 193 \Angst\ (left) and \stereo-A 195 \Angst\ (right) at 08:25:46 UT with the \hsi 6-12 and 25-50~keV contours of this flare.

\begin{figure*}[htb!]
\begin{center}
\includegraphics[width=0.8\textwidth]{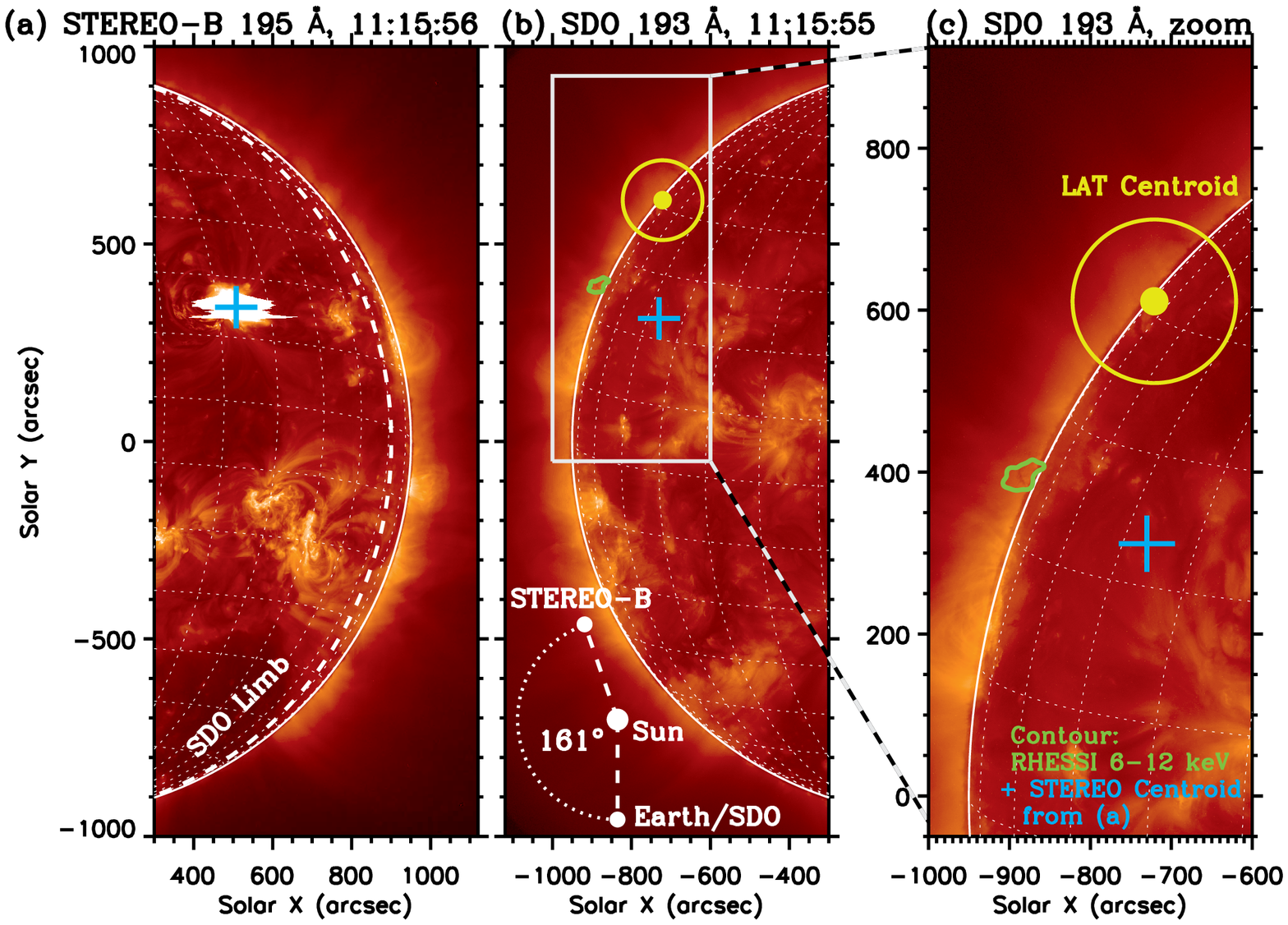}
  \caption{Localization of the Sep14 flare. Images near the time of the flare peak are shown for \stereo-B 195 \Angst\ (a), \sdo 193  \Angst\ (b) and an enlargement of the \sdo image (c) marked by the white rectangle in (b).  The yellow circle in (b) and (c) shows the 68\% error
circle for the \Fermi-LAT emission centroid. The time range of the \Fermi-LAT data is from 11:02:00--11:20:00 UT. The green contour in (b) shows the 6--12 keV \hsi source. The blue cross in (a) marks the {\it STEREO} flare ribbon centroid, whose projected position as seen by \sdo\ is shown in (b) and (c) with the same symbol. The white-dashed line in (a) represents the solar limb as seen by \sdo.}
\label{fig:SOL20140901Loc}
\end{center}
\end{figure*}

The \Fermi-LAT $>$100 MeV emission centroid of Sep14 is located at heliocentric
coordinates [-720$''$,610$''$] with a 68\% error radius of 100$''$. \hsi imaging
shows a 6--12 keV source located above the visible limb slightly offset from the \Fermi-LAT centroid, both shown in Figure~\ref{fig:SOL20140901Loc}. If
the \hsi source is the loop-top of the behind-the-limb flare then the minimum
height needed for this source to be visible from $\sim$40$^{\circ} $
behind-the-limb would be $\sim$10$^{10}$ cm. HXR loop-top emission from a flare located $\sim$40$^{\circ} $ behind the limb has been detected before by
\hsi~\citep{KruckerS_Coronal60keV_2007ApJ...669L..49K}. The \Fermi-LAT measured 17
photons with energies $>$1 GeV; 15 of these (including a 3.5 GeV photon with an arrival time of 11:16:01 UT) arrived during the first 20 minutes of \Fermi-LAT detection.

\section{Spectral analysis}\label{sec:analysis}

\subsection{Gamma-ray spectra}
\label{sec:gamma}

We performed an unbinned likelihood analysis of the \Fermi-LAT data with the \texttt{gtlike} program distributed with the \Fermi \texttt{ScienceTools}\footnote{We used the version 10-01-00 available from the \Fermi Science Support Center \url{http://fermi.gsfc.nasa.gov/ssc/}}. We selected Pass 8 Source class events from a 10$^{\circ}$ circular region centered on the Sun and within 100\de from the local zenith (to reduce contamination from the Earth limb). 

We fit three models to the \Fermi-LAT gamma-ray spectral data. The first two, a pure
power law (PL) and a power-law with an exponential cut-off (PLEXP) are
phenomenological functions that may describe bremsstrahlung emission from
relativistic electrons. The third model uses templates based on a detailed study
of the gamma rays produced from decay of pions originating from
accelerated protons with an isotropic pitch angle distribution in a thick-target
model~\citep[updated from][]{murp87}. 

We rely on the likelihood ratio test and the associated test statistic TS~\citep[]{Mattox:96} to estimate the significance of the detection. The TS is defined as twice the increment of the logarithm of the likelihood obtained by fitting the data with the source and background model components simultaneously. Because the null hypothesis (i.e. the model without an additional source) is the same for the PL and PLEXP models, the increment of the TS ($\Delta$TS=TS$_{\rm PLEXP}$-TS$_{\rm PL}$) is equivalent to the corresponding difference of maximum likelihoods computed between the two models. Note that the significance in $\sigma$ can be roughly approximated as $\sqrt{\rm TS}$ for 2 degrees of freedom.


\begin{deluxetable*}{ccccccc}[ht]
  \tablewidth{\textwidth}
  \tablecaption{\Fermi-LAT Spectral Analysis of the Solar flares considered in this work}
  \tablehead{\colhead{Time Interval}  & \colhead{TS$_{PL}$} & \colhead{$\Delta$TS$^{a}$}  & \colhead{Photon Index$^{b}$} & \colhead{Cutoff Energy$^{c}$} & \colhead{Flux$^{d}$} & \colhead{Proton Index} \\
    \colhead{(UT)}  & \colhead{}  & \colhead{} & \colhead{}& \colhead{(MeV)} & \colhead{($\times$10$^{-5}$ ph cm$^{-2}$ s$^{-1}$)} & \colhead{}}
  \startdata
  \multicolumn{7}{c}{SOL2013-10-11 }\\ 
  \hline
  07:10:00--07:15:00 & 20 & 3 & $-$2.0$\pm$0.4 & - & 0.9$\pm$0.1  & - \\
07:15:00--07:18:00 & 487 & 43 & $-$0.4$\pm$0.4 & 130$\pm$30 & 17$\pm$1 & 4.5$\pm$0.4 \\
07:18:00--07:20:00 & 846 & 77 & $-$0.1$\pm$0.4 & 112$\pm$21 & 49$\pm$2 & 4.3$\pm$0.2 \\
07:20:00--07:22:00 & 776 & 76 & $-$0.3$\pm$0.3 & 128$\pm$23 & 46$\pm$2 & 4.4$\pm$0.3 \\
07:22:00--07:24:00 & 677 & 58 & $-$0.2$\pm$0.4 & 153$\pm$31 & 29$\pm$2 & 3.7$\pm$0.2 \\
07:24:00--07:26:00 & 345 & 25 & $-$0.9$\pm$0.4 & 208$\pm$61 & 21$\pm$2 & 3.9$\pm$0.3 \\
07:26:00--07:28:00 & 219 & 34 & 0.6$\pm$0.7 & 86$\pm$22 & 16$\pm$1 & 4.5$\pm$0.4 \\
07:28:00--07:40:00 & 283 & 29 & 0.1$\pm$0.7 & 86$\pm$26 & 7$\pm$1 & 5.3$\pm$0.4 \\
\hline\hline\\
\multicolumn{7}{c}{ SOL2014-01-06 }\\
 \hline
07:55 -- 08:15 & 67 & 20  & $-$2.4$\pm$0.2 & - & 0.6$\pm$0.1 &  -\\ \\
\hline\hline\\
\multicolumn{7}{c}{ SOL2014-09-01 }\\
\hline
11:02:00--11:04:00 & 42 & 9 & $-$2.5$\pm$0.3 & - & 11$\pm$3 & - \\
11:04:00--11:06:00 & 321 & 41 & $-$0.2$\pm$0.5 & 117$\pm$27 & 117$\pm$10 & 4.7$\pm$0.4 \\
11:06:00--11:08:00 & 1070 & 120 & $-$0.5$\pm$0.3 & 116$\pm$15 & 360$\pm$17 & 5.2$\pm$0.2 \\
11:08:00--11:10:00 & 1549 & 126 & $-$0.9$\pm$0.2 & 158$\pm$20 & 477$\pm$19 & 4.9$\pm$0.2 \\
11:10:00--11:12:00 & 2549 & 115 & $-$1.2$\pm$0.2 & 194$\pm$28 & 565$\pm$21 & 4.8$\pm$0.2 \\
11:12:00--11:14:00 & 6394 & 157 & $-$0.8$\pm$0.1 & 167$\pm$10 & 522$\pm$20 & 4.4$\pm$0.1 \\
11:14:00--11:16:00 & 2430 & 159 & $-$0.6$\pm$0.2 & 148$\pm$17 & 540$\pm$22 & 4.5$\pm$0.2 \\
11:16:00--11:18:00 & 1069 & 32 & $-$1.9$\pm$0.2 & 488$\pm$90 & 465$\pm$23 & 4.2$\pm$0.2 \\
11:18:00--11:20:00 & 1047 & 49 & $-$1.0$\pm$0.3 & 182$\pm$37 & 396$\pm$27 & 4.5$\pm$0.2 \\
12:29:00--12:34:00 & 56 & 12 & $-$2.2$\pm$0.2 & - & 3$\pm$1 & - \\
12:34:00--12:39:00 & 133 & 3 & $-$2.3$\pm$0.2 & - & 3$\pm$1 & - \\
12:39:00--12:44:00 & 130 & 5 & $-$2.2$\pm$0.2 & - & 2$\pm$1 & - \\
12:44:00--12:49:00 & 97 & 11 & $-$2.3$\pm$0.2 & - & 2$\pm$1 & - \\
12:49:00--12:54:00 & 84 & 3 & $-$2.4$\pm$0.2 & - & 2$\pm$1 & - \\
\enddata
\label{tab:SpectralAnalysis}
\tablenotetext{a}{$\Delta$TS=TS$_{\rm PLEXP}$-TS$_{\rm PL}$}
\tablenotetext{b}{Photon index from best-fit model. The PL is defined as $\frac{dN(E)}{dE} = N_0E^{\Gamma}$ and the PLEXP as $\frac{dN(E)}{dE} = N_0E^{\Gamma}\exp(-\frac{E}{E_{c}})$ where $E_c$ is the cutoff energy.}
\tablenotetext{c}{From the fit with the PLEXP model.}
\tablenotetext{d}{Integrated flux between 100\,MeV and 10\,GeV calculated for
the best-fit model.}

\end{deluxetable*}

In Table~\ref{tab:SpectralAnalysis} we list the TS$_{\rm PL}$, $\Delta{\rm TS}$, $\Gamma$ the photon index for the best-fit model (PL when $\Delta{\rm TS}<25$ or PLEXP when $\Delta{\rm TS}\geq25$) and PLEXP cut-off energy. For several intervals  $\Delta{\rm TS}>$25, indicating that PLEXP provides a significantly better fit than PL. For these intervals we fit the pion-decay models to the data to determine the best proton spectral index following the same procedure described in~\citet{0004-637X-789-1-20}. In particular,  we performed a series of fits with the pion-decay template models calculated for a range of proton spectral indices. We then fit the resulting profile of the log-likelihood with a parabolic function of the proton index. The minimum gives the most likely index for the pion-decay model. Note that the TS values for PLEXP and pion-decay fits cannot be directly compared~\citep{wilks1938} because they are not nested models. However the PLEXP approximates the shape of the pion-decay spectrum; thus we expect the pion-decay models to provide a similarly acceptable fit.

The main contributions to the systematic uncertainties are uncertainties in the effective area; in the energy range 100~MeV to $\sim$100~GeV these are of the order of $\pm$5\%.This uncertainty applies directly to the flux values and from our previous studies on LAT-detected solar flares~\citep{0004-637X-789-1-20} we find that the systematic uncertainties on the cut-off energy and photon index are also of the order of $\pm$5\%\footnote{A more detailed description of the systematic uncertainties tied to the effective area of the LAT can be found here \url{http://fermi.gsfc.nasa.gov/ssc/data/analysis/LAT_caveats.html}.}.

\begin{deluxetable*}{ccccccccc}[ht]
  \tablewidth{\textwidth}
\tablecaption{Comparison between behind-the-limb and on-disk flare quantities}
\tablewidth{0pt}
\tablehead{
  \colhead{Date} & \colhead{GOES$^{a}$} & \colhead{CME speed$^{b}$}  & \colhead{AR } & \colhead{Duration} & \colhead{Peak Flux$^{c}$}  & \colhead{E$_{\gamma}>$100 MeV$^d$} & \colhead{Proton$^{e}$} & \colhead{E$_p$ $>$500 MeV$^{f}$} \\
\colhead{(UTC)}  & \colhead{class} & \colhead{(km s$^{-1}$)}  & \colhead{position} & \colhead{(minutes)}&  \colhead{(10$^{-5}$ ph cm$^{-2}$ s$^{-1}$)} &\colhead{(ergs)} & \colhead{index} & \colhead{(ergs)} }
  \startdata
  2013-10-11 & M4.9 & 1200 & N21E103 & 30  & 49$\pm$2 & 1.5$\times$10$^{23}$ & 4.3$\pm$0.1  & 9.8$\times$10$^{24}$  \\
  2014-01-06 & X3.5 & 1400 & S8W110  & 20  & 0.8$\pm$0.1 & 4.2$\times$10$^{21}$ & 5.3$\pm$0.4$^{g}$ & 3.5$\times$10$^{23}$ \\
  2014-09-01 & X2.4 & 2000 & N14E126 & 113 & 565$\pm$14 & 1.4$\times$10$^{24}$ & 4.7$\pm$0.1  & 7.0$\times$10$^{25}$ \\
  \hline \hline
  \multicolumn{9}{c}{On-disk flares}\\
  \hline
  2011-03-07 & M3.7 & 2125 & N30W48  & 798 & 3$\pm$1 &  5.1$\times$10$^{23}$& 4.7$\pm$0.2 & 3.6$\times$10$^{25}$  \\
  2011-06-07 & M2.5 & 1255 & S22W53  & 38 & 3$\pm$1 & 3.2$\times$10$^{22}$ & 5.0 $\pm$0.3 & 2.5$\times$10$^{24}$ \\
  2012-03-07I$^{h}$ & X5.4 & 2684 & N16E30  & 45  & 417$\pm$13 &  3.9$\times$10$^{24}$ & 3.90$\pm$ 0.02  & 2.1$\times$10$^{26}$\\
  2012-03-07E$^{i}$ & X5.4 & 2684 & N16E30  & 1068 & 97$\pm$2 &  1.4$\times$10$^{25}$ & 4.3$\pm$0.1 & 9.0$\times$10$^{26}$  \\
    \enddata
    \label{tab:BTLQuantities}
    \tablenotetext{a}{GOES class for behind-the-limb flares is estimated based on the \stereo 195 \Angst\ flux.}
    \tablenotetext{b}{Speed is the linear speed reported by the {\it LASCO} CME catalog.}
    \tablenotetext{c}{For photon energies $>$100 MeV.}
    \tablenotetext{d}{Total energy released in $>$100 MeV gamma rays  integrated over the time interval when $\Delta$TS$>$25.}
    \tablenotetext{e}{Proton index in the same time interval as (d).}
    \tablenotetext{f}{Total energy released by protons with E$>$500 MeV estimated over the same interval as in (d). Values may be underestimated for flares with centroids at heliocentric angles $>$75$^{\circ}$.}
    \tablenotetext{g}{The $\Delta$TS value for this flare is 20, so the improvement over a power law is marginal.}
    \tablenotetext{h}{Impulsive phase of the flare, \Fermi-LAT detection from 00:38:52 - 01:23:52 UT. Note that the peak of the GOES X-ray flare occurred $\sim$6 minutes prior to \Fermi orbital sunrise.}
    \tablenotetext{i}{Extended phase of the flare, starting from 02:27 UT.}
\end{deluxetable*}

\begin{deluxetable}{lc}[ht]
   \tablewidth{0.47\textwidth}
   \tablecaption{Best-Fitting Spectral Parameters of \Fermi-LAT and \Fermi-GBM data}
   \tablehead{
     \colhead{Parameter} & \colhead{Value}}
  \startdata
  \multicolumn{2}{c}{SOL2013-10-11}\\
   \hline\\
   \multicolumn{2}{c}  {Best fit model: PL1+PION} \\
   Power-law (1) index  & 3.2$\pm$0.1 \\   
   Proton Index         & 4.1$\pm$0.1\\
   \multicolumn{2}{c}  {PL1+PL2$\times$EXP+PION} \\
   Power-law (1) index  & 3.4$\pm$0.06 \\
   Power-law (2) index  & $-$0.9$\pm$0.3\\
   Cutoff energy (MeV) & 0.8$\pm$0.1 \\
   Proton Index         & 4.1$\pm$0.1\\
  \hline\\
   \multicolumn{2}{c}{SOL2014-09-01}\\  
    \hline\\
    \multicolumn{2}{c}  {Best fit model: PL1$\times$EXP+PION} \\
    Power-law (1) index & 2.06$\pm$0.01\\
    Cutoff energy (MeV) & 90$\pm$7 \\
    Proton Index & 4.4$\pm$0.1\\
   \multicolumn{2}{c}  {(PL1+PL2)$\times$EXP+PION} \\
    Power-law (1) index & 2.18$\pm$0.01\\
    Power-law (2) index & 1.4$\pm$0.3 \\
    Cutoff energy (MeV) & 10$\pm$0.1 \\
    Proton Index & 4.4$\pm$0.1\\
     \enddata
    \label{tab:SpectralFits}
    \tablenotetext{}{Model parameters for both the Oct13 and Sep14 flares. For the Oct13 flare we integrated between 7:10 and 7:35 UT and for the Sep14 we integrated from 11:02 and 11:20 UT. The parameters of the best-fit model are compared with those of a more complex model with double power laws at low energy (whose improvement is not statistically significant).}
   
\end{deluxetable}

\subsubsection{Comparison of BTL and Disk flare Characteristics}

We  compare the characteristics of the $>100$ MeV emission associated with the
three BTL flares with three disk flares with similar \goes classifications and
temporally extended emissions and are described in \citet{2014ApJ...787...15A} and \citet{0004-637X-789-1-20}. In addition
to spectral parameters we also compare the total $>100$ MeV emitted
energies and the total energy released by protons with energy $>500$ MeV needed to
produce the detected gamma-ray emission, as based on the templates of~\citet[updated from][]{murp87}. We
present these numbers along with the date, estimated \goes class, CME speed, AR 
position and the \Fermi-LAT detection duration in Table~\ref{tab:BTLQuantities}. The proton indexes are very similar whereas the on-disk flares appear to have more energy; this is most likely because we observed the on-disk flares over longer time scales.  Peak fluxes and the total energy released by protons with E$>$500 MeV for Sep14 and the impulsive phase of SOL2012-03-07 are comparable.

\begin{figure}[ht]
  \begin{center}
    \includegraphics[width=0.52\textwidth]{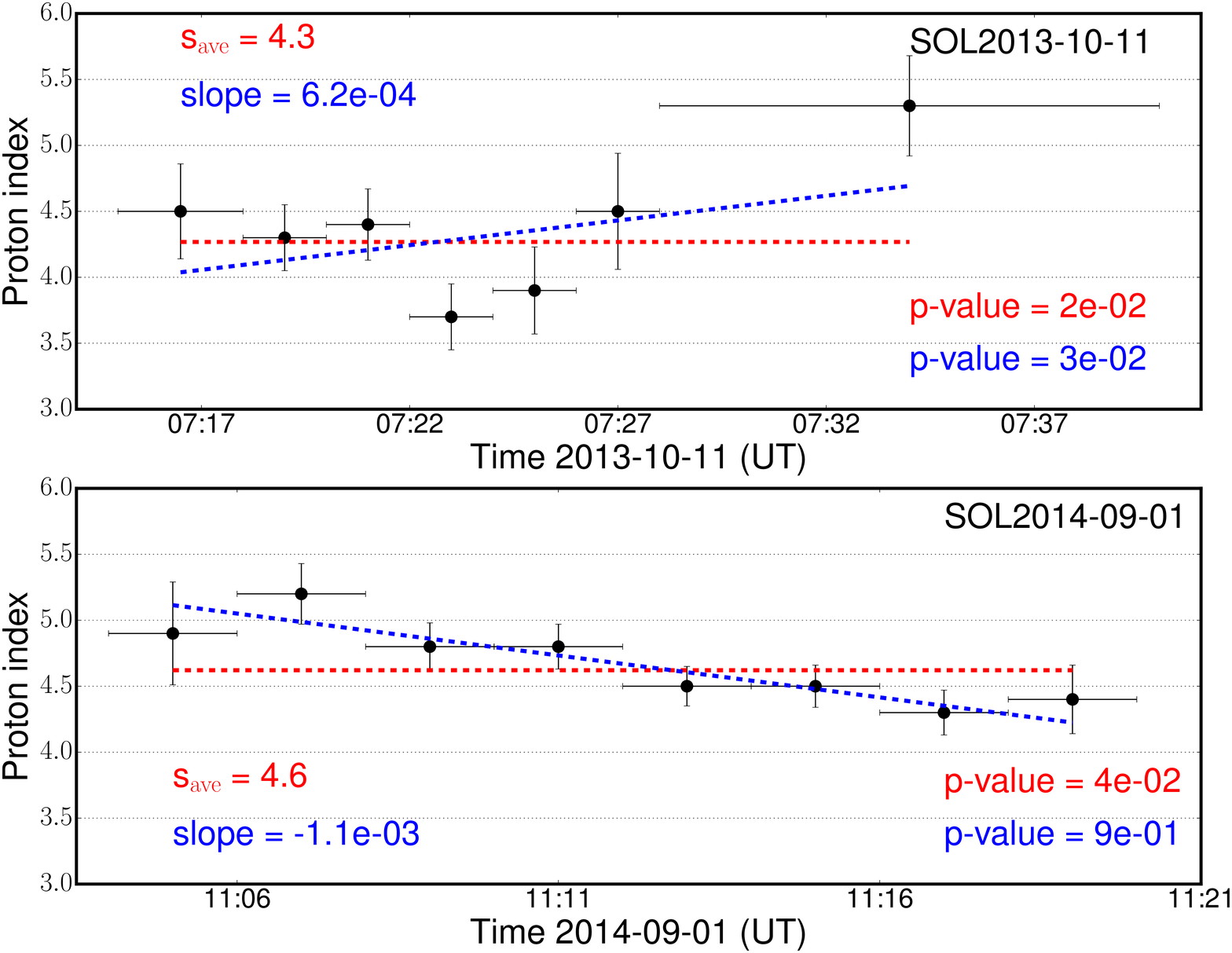}
\caption{Proton index inferred from the fit of the gamma-ray emission with the pion decay templates as a function of time for Oct13 (top panel) and Sep14 (bottom panel). The red-dashed lines represent a fit with a constant and the blue dashed lines a linear fit. Fit parameter values and p-values are indicated in each panel.}
\label{fig:OnDisk_BTL_protonIndex}
\end{center}
\end{figure}

Figure~\ref{fig:OnDisk_BTL_protonIndex} shows the proton index as a function of time for
the Oct13 and Sep14 BTL flares. We fit a constant (red dashed line) and a first-degree polynomial (blue dashed line) to the data. In the case of the constant fit we list the best-fit
value in the upper left corner, whereas for the straight-line fit
we list the value of the slope. The temporal variation over tens of minutes is not sufficient to conclude whether a softening or hardening is present for these BTL flares.
In~\citet{0004-637X-789-1-20} we found that for the on-disk flare SOL2012-03-07 the spectrum softened with a time scale of a few hours.
\begin{figure}[htb!]
\begin{center}
  \includegraphics[width=0.49\textwidth]{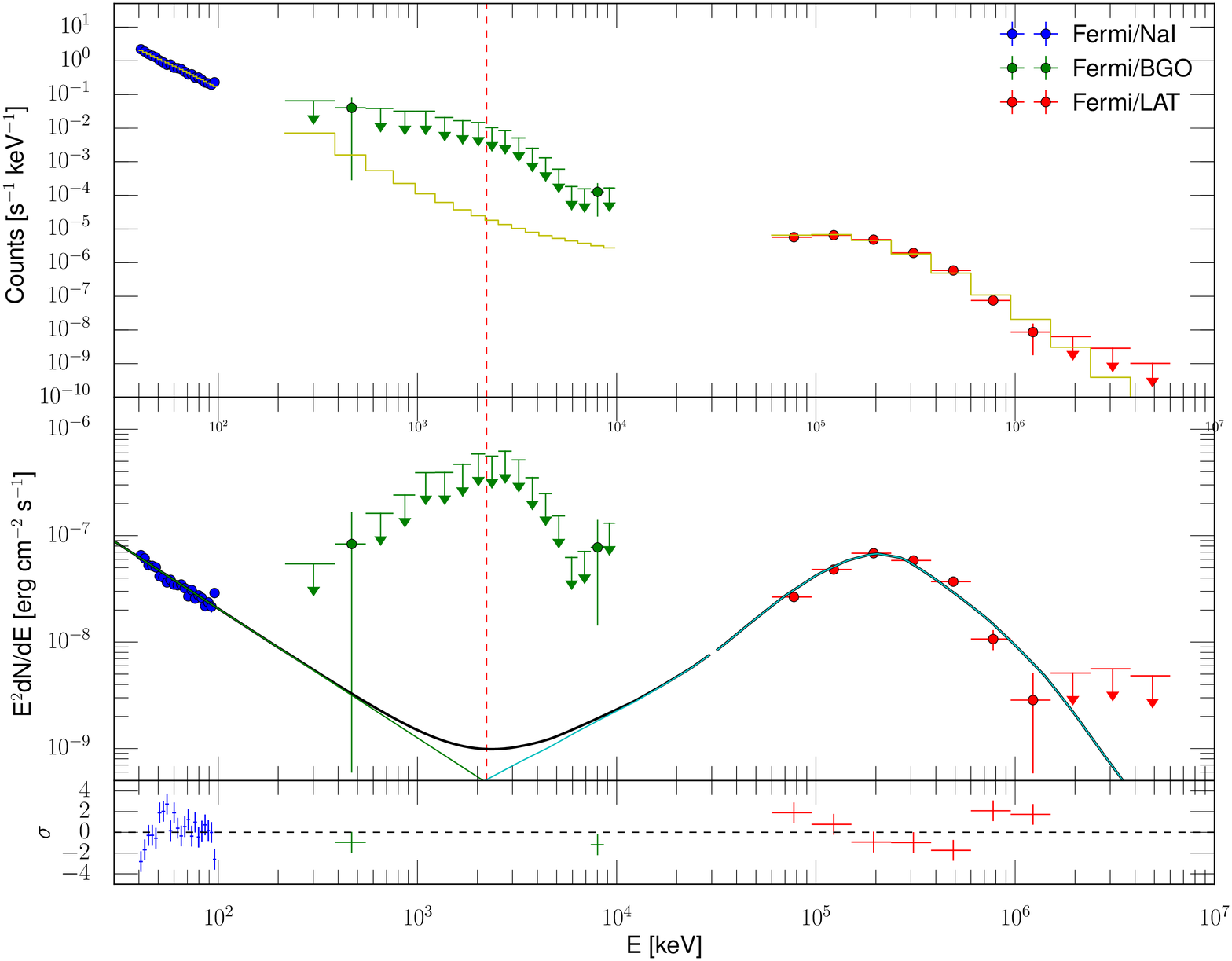}
  
\caption{Combined \Fermi-GBM/LAT count spectra (top panel) and spectral energy distribution (middle panel) for the Oct13 flare integrated between 7:10 and 7:35 UT and the lower panel shows the residuals of the fit.
The model that best-fits the data is a power-law that dominates the low-energy spectrum, and a pion-decay model, which describes the \Fermi-LAT spectrum. The neutron capture line (at 2.223 MeV, highlighted by the red vertical dashed line) is not statistically significant, and neither is an additional low energy power law with exponential cut off ($\sim$2.5$\sigma$). The parameters of the fits are listed in Table~\ref{tab:SpectralFits}.}
\label{fig:SEDSOL20131011}
\end{center}
\end{figure}

\begin{figure}[htb!]
\begin{center}
  \includegraphics[width=0.49\textwidth]{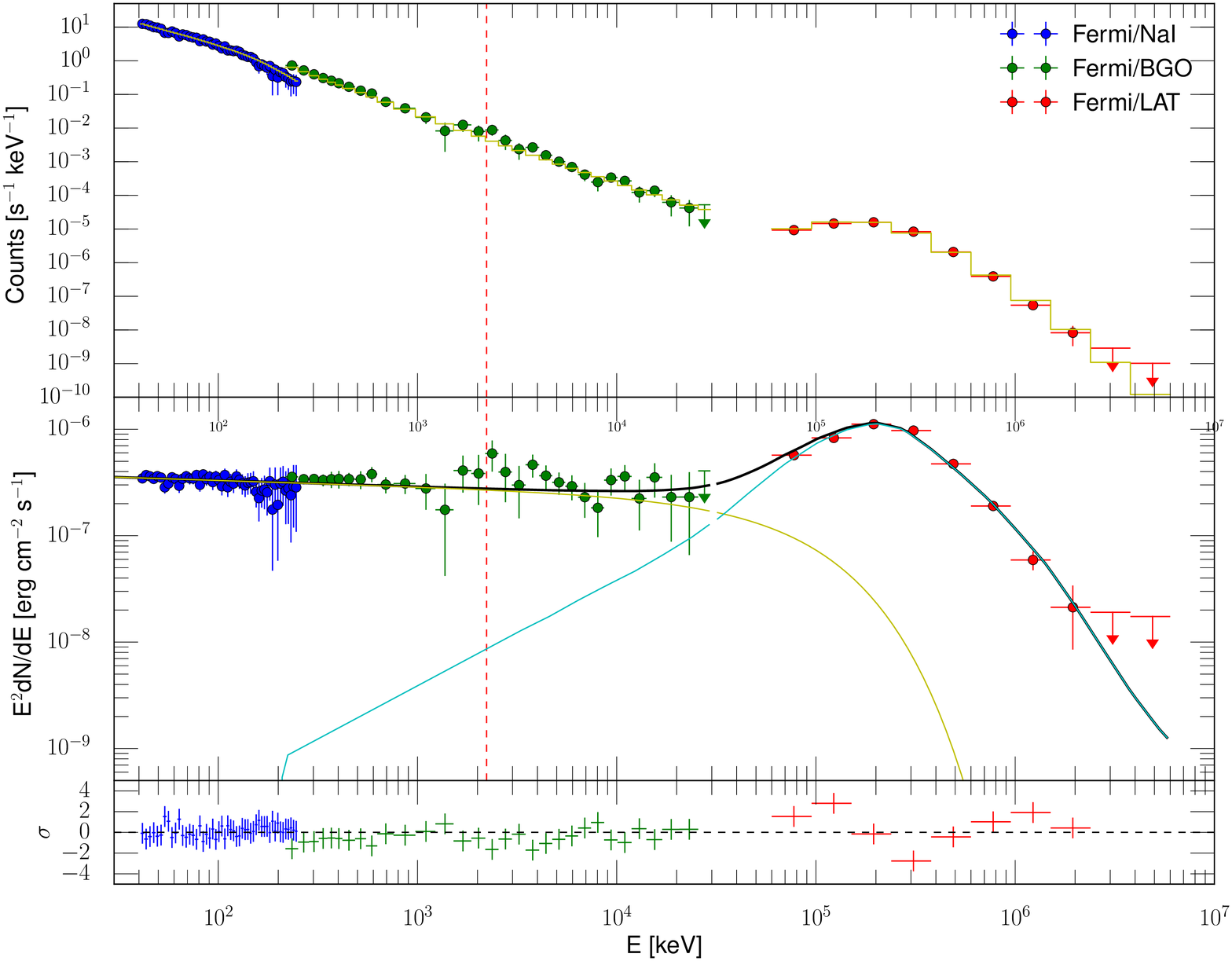}
  \caption{Combined \Fermi-GBM/LAT count spectra (top panel) and  SED (middle panel) for the Sep14 flare integrated between 11:02 and 11:20 UT and the lower panel shows the residuals of the fit. 
The best-fit model is a power law with exponential decay at high energy to describe the emission from 30 keV to $\sim$10 MeV and a pion-decay model to describe the \Fermi-LAT spectrum.
The neutron capture line (at 2.223 MeV, highlighted by the red vertical dashed line) is not statistically significant ($\sim$2$\sigma$) and neither is an additional power law at low energy $\sim$2$\sigma$. The fit parameters are listed in Table~\ref{tab:SpectralFits}.}
\label{fig:SEDSOL20140901}
\end{center}
\end{figure}

\subsection{X-ray spectra}
\label{Xray}
As described above we have HXR data from three instruments: \hsi, \Fermi-GBM
and \konus. \konus was in waiting mode during the
Jan14 and Sep14 flares; therefore these events were detected
in only three broad energy channels, namely $\sim$20--78, 78--310 and 310--1180 keV with
2.944 s time resolution. 
\hsi provides only a limited coverage of these flares. For the Oct13 flare
there is some overlap with \Fermi-GBM during the rise of the impulsive phase. As
shown in~\citet{2015ApJ...805L..15P} the HXR spectra of \hsi and \Fermi-GBM agree
well. \Fermi-GBM was in the SAA and \hsi was in spacecraft night during the impulsive phase of the Jan14
flare and detected the flare during the decay phase in only two low-energy
channels. Thus,  we do not have sufficient data for a spectral fit. For the
Sep14 flare \hsi gives information only on the decay phase of the flares. In the
following we show results from analysis of \Fermi-GBM data for the Oct13 and Sep14 flares.

\subsection{Combined Spectroscopic Studies}
\label{combined}
For both Oct13 and Sep14 we performed a combined \Fermi-GBM/LAT fit using the \texttt{XSPEC} package \citep{XSPEC}.
The spectral fits were done by minimizing the \PGSTAT. \PGSTAT is a profile likelihood statistic that takes into account Poisson error on the total count spectrum and Gaussian error on the background.
In order to obtain the background-subtracted spectra of the \Fermi-GBM data for Sep14 we used both \Fermi-GBM-NaI and \Fermi-GBM-BGO spectra accumulated before the flare (from 10:54 UT to 10:57 UT),  and after the flare (from 11:42 - 11:50 UT).
For Oct13, because there was a minor on-disk flare whose onset time was 7:01 UT, we used the procedure described in \citet{2015ApJ...805L..15P}, which consists of using the background estimation tool developed in \citet{2012SPIE.8443E..3BF} with an additional 5\% systematic error.
For the LAT, we follow the procedure used in \citet{0004-637X-789-1-20} and \citet{2015ApJ...805L..15P} that consists of deriving the background spectrum directly from the model of the background used in the standard LAT likelihood analysis (using first \texttt{gtlike} and then \texttt{gtbkg}), and obtaining  the response of the LAT using \texttt{gtrspgen} (all of these tools are available in the \Fermi \texttt{ScienceTools}).
The combined spectral energy distribution (SED) derived from the \Fermi-GBM and \Fermi-LAT data extending from 30~keV to 10~GeV for these flares is shown in
Figures~\ref{fig:SEDSOL20131011} and \ref{fig:SEDSOL20140901}, respectively, and the parameters for the spectral fit for both solar flares are listed in Table~\ref{tab:SpectralFits}.

The Oct13 flare had a very weak signal in the BGO, and the best-fit model (M$_{0}$) consisted of a single power law and the pion-decay templates to describe the bremsstrahlung and $>$60 MeV emission detected by the \Fermi-GBM and \Fermi-LAT, respectively.
For the Sep14 flare, the best-fit model (M$_{0}$) consisted of a single power law with an exponential cut off at high energies to describe the bremsstrahlung emission, and the pion-decay templates to describe the $>$60 MeV emission detected by the \Fermi-LAT, similar to what was done for the 12 June 2010 \Fermi flare~\citep{2012ApJ...745..144A}.
We also  tested the statistical significance of an extra power law with a high-energy cut off (model M$_{1}$).
To this end we performed Monte Carlo simulations (using XSPEC \texttt{fakeit}) by generating a spectrum described by the best fit model (M$_{0}$). 
We then re-optimized the parameters of M$_{0}$ by fitting it to the simulated data. We finally compared the improvement of \PGSTAT when fitting the simulated data with the model M$_{1}$.
The improvement of \PGSTAT for the Oct13 and Sep14 flares is $\Delta_{\PGSTAT}\approx$408 and $\approx$109, respectively. 
Despite these high values of $\Delta_{\PGSTAT}$, dedicated Monte Carlo simulations indicate that they correspond to significances of 2.5 and 2.0 $\sigma$, respectively.

We then checked whether the addition of the 2.223 MeV line (on top of model M$_{0}$) significantly improved the fit. For the Oct13 flare the improvement is negligible ($\Delta_{\PGSTAT}$$\approx$0), while for the Sep14 the improvement is ($\Delta_{\PGSTAT}$$\approx$25), corresponding to a significance of 2.0 $\sigma$, estimated from Monte Carlo simulations.
We tested whether adding nuclear de-excitation narrow lines and continua provided an improvement to the fit for both flares and found that it was not significant. The nuclear line and continua templates used here are based on a detailed study of the nuclear gamma-ray production from accelerated ion interactions with elements found in the solar atmosphere~\citep{murp09}.
\begin{figure}[htb!]
\begin{center}
\includegraphics[width=0.46\textwidth]{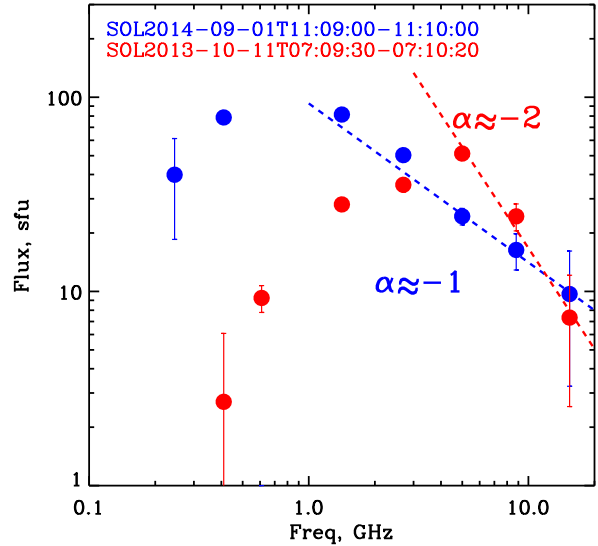}
\caption{Radio spectra during the peaks of the HXR and radio time profiles (07:09:30--07:10:20 UT) for the Oct13 flare in red and for the Sep14 flare in blue (11:09:00--11:10:00 UT). The microwave photon spectral index $\alpha$, found by fitting the optically thin part of the spectra for each flare, is also shown.}
\label{fig:RadioSpecOct13}
\end{center}
\end{figure}

\subsection{Radio Spectra} 
\label{sec:radio}

The radio spectra obtained at the peaks of the Oct13 and Sep14 events are shown in  Figure~\ref{fig:RadioSpecOct13}. The spectrum was obtained by using the RSTN 1-s resolution fluxes. The spectra were background subtracted and the fluxes were integrated within the selected time intervals for each frequency. For each spectrum we detected the frequency where the maximum flux was observed. The spectra of the Oct13 and Sep14 flares are similar to those
expected from the gyro-synchrotron (GS) mechanism, with peaks that separate the optically thin and thick parts  \citep[for GS spectrum properties see, for example,][]{1985ARA&A..23..169D}. The values for the microwave  photon spectral index, $\alpha$, for both flares are shown in Figure~\ref{fig:RadioSpecOct13}. Unfortunately, the RSTN data for the time period of the Jan14 flare are not sufficient to obtain radio spectra.

\section{Solar Energetic Particles (SEP)}
\label{sec:Solar Energetic Particles}
\begin{figure}[ht]
\begin{center}
 \includegraphics[width=0.48\textwidth]{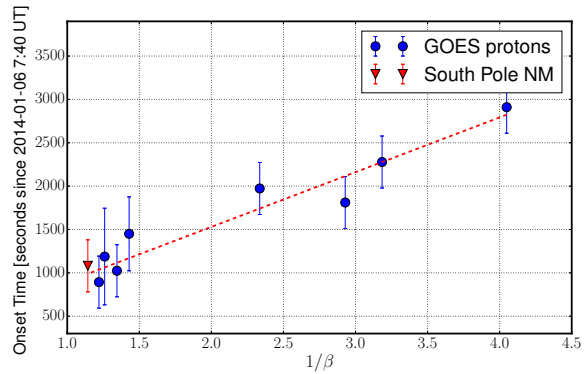}
  \caption{Onset times as a function of 1/$\beta$ for the GOES SEP protons associated with Jan14. The Y-axis gives the time in seconds since  2014-01-06 07:40 UT (assumed start of the flare as detected by \stereo-B).  Assuming scatter-free propagation for the first arriving particles. The red dashed line represents the fit using Equation~\ref{eq:seponsetimes}, obtaining the pathlength $\zeta=1.3\pm0.2$ AU and $T_1=$07:55$\pm$00:05 UT.}
\label{fig:SOL20140106OnsetTimes}
\end{center}
\end{figure}

Interpretation of the origin of SEPs is more complicated than studying photons,
since they move with different velocities along curved magnetic field lines and
are detected only when the field lines  are connected
to the instrument. Moreover, they may be scattered by turbulence with energy-dependent mean free path. The \goes proton fluxes did not show a significant increase for the Oct13 event (which occurred behind the eastern
limb). However, \stereo-B, positioned behind the limb and with better magnetic
connectivity, detected an increase of SEP proton intensity starting roughly 2 hours
after the detection of an electromagnetic signal from this flare. Whereas the Sep14 flare, which was located roughly 40$^{\circ}$ behind the eastern limb, was associated with an increase in the proton fluxes detected both by \stereo-B (which had a front view of the flare) and \goes (starting roughly 9 hours after the flare).

The Jan14 flare, originating in an AR 20$^{\circ}$ behind the western limb, produced no significant increase in the proton flux at \stereo,  had better magnetic connectivity to the Earth and  was associated with a very strong
SEP event with neutrons detected on ground by the South Pole neutron
monitors.

The arrival time~$T_2$~\citep{JGRA:JGRA8939} at the Earth of  a
particle with velocity $v=\beta c$ released from the Sun at time $T_1$,  is:
\begin{equation}
  T_2(\frac{1}{\beta}) = T_1 + \frac{\zeta}{c}\times\frac{1}{\beta}
  \label{eq:seponsetimes}
\end{equation}
where $\zeta$ is the distance traveled by the SEP from the acceleration region
to the Earth. Assuming scatter-free propagation for the first-arriving
particles, $\zeta$ will be independent of particle energy or velocity and will be equal to  the path length along a Parker spiral magnetic field line from the Sun to the Earth (usually assumed to be 1.2 AU). If all the
particles are accelerated at the same place and time, then the intercept
of the line fitting $T_2$ to $1/\beta$ would give $T_1$, the time at which SEPs are released and is known as the Solar Particle Release (SPR) time. The slope of this line, $\zeta/c$ will be of the order of 600~s.

Since we do not have high energy resolution SEP data we  obtain the
SEP onset times using  two methods depending on the energy range of the
particles. For the GOES SEPs with energies $<$200~MeV we apply a median filter to
the intensity profile and define the onset time as when the profile reached 5\% of the maximum.
For the higher-energy SEPs we apply a Fast Fourier Transform (FFT) filter to the
intensity profile and evaluate the time when the second derivative is maximum as
the onset time. To estimate of the uncertainty on the onset time
we performed a scan over a series of values for the median and FFT filter windows
and take the difference in onset times as the error. 

The onset times as a function of $\frac{1}{\beta}$ for the \goes SEP protons associated with Jan14 are shown in Figure~\ref{fig:SOL20140106OnsetTimes}.  Fitting the data to Equation~\ref{eq:seponsetimes}, we estimate the SPR time to be $T_1 = $07:55 $\pm$ 0:05 UT and path length of $\zeta\sim 1.3\pm 0.2$ A.U.

 The value for $T_1$ is consistent with the estimated SPR time, 07:47 UT, reported by~\citet{2014ApJ...790L..13T} but is somewhat later than the start of the flare activity at the Sun. Since \Fermi was in the SAA and \hsi was in orbital night, the most accurate time of the initiation of the flare was obtained from \konus and \stereo data. \konus detected emission in the 21--78 keV
energy range starting at 07:43 UT, implying an emission time at the Sun of 07:35
UT, roughly 12 minutes prior to the SPR time. Assuming that the CME was released at this latter time the acceleration of SEPs started when the
CME was at a height of  1.5 \Rs$(v_{\rm CME}/$1400 km s$^{-1})$. 

Unfortunately the high-energy SEP data were not sufficient to estimate the SPR time for the other two BTL flares occurred on Oct13 and Sep14.

\section{Summary and Discussion}\label{sec:discussion}

We have presented the analysis of \Fermi-LAT  data from  three BTL solar flares,
whose location is
determined by \stereo observations. We also presented analyses of
data from \hsi, \Fermi-GBM, \konus in the HXR range, \sdo in EUV,  RSTN in
radio wavelengths, some SEP observations, GOES and South Pole Neutron Monitors. We found several results:

\begin{enumerate}

\item

The HXR LCs measured by the three instruments (in overlapping energy
ranges) are in good agreement with each other  and with the high-frequency radio
LCs. The \Fermi-LAT emission, in general, commences several minutes after the HXRs, peaks at a later time and lasts longer.

\item

We are able to obtain \hsi images in soft HXRs  for flares just over the limb,
indicating a loop-top emission from loops with heights above the
photosphere ranging  from $0.2\times 10^{10}$ to $2\times 10^{10}$ cm. \sdo detected UV emission is detected for Oct13, and possibly Jan14, but not the Sep14 flare. Jan14 flare
was too weak to permit localization of the gamma-ray source but localizations
of the \Fermi-LAT emission with Pass 8 data for the other two flares situates the centroid locations near the solar limb and
about 50$\arcsec$ (Oct13) and 300$\arcsec$ (Sep14) from the \hsi centroid. 

\item

The LCs (available only for Oct13 and Sep14 flares) 
show good agreement between the HXR spectra measured by
\Fermi-GBM, \hsi (whenever contemporaneous data are
available) and  \konus below a few 100 keV. The
\Fermi-LAT spectra can be fitted with a relatively hard PLEXP phenomenological model
(possibly due to relativistic electron bremsstrahlung) or to
a pion decay model with proton indexes around 4. We note that production of
photons of about $\sim$3 GeV detected here requires $>15$ GeV protons or $>6$ GeV electrons. These energies are greater than any other reported for solar particles. 

\item

We compare several spectral properties of $>100$ MeV radiation from the BTL
flares with those of three on-disk flares previously analyzed. As evident from 
Table~\ref{tab:BTLQuantities} and 
Figure~\ref{fig:OnDisk_BTL_protonIndex}, in general, these flares have 
similar characteristics. Similar \goes class flares have similar peak
photon fluxes, surprisingly, regardless of limb occultation. The higher total  gamma-ray energy of on-disk flares is most likely results from their having been observed over longer timescales.  Thus, the underlying acceleration and emission processes are most likely similar, but the transport paths of the radiating particles (presumably protons) from the acceleration to the emission site must be different. That is, for BTL flares, protons must travel greater distances to land on the front side of the Sun to produce the detected gamma-rays.

\item 

We find that the proton index  does not vary significantly during the 30-minute observation time windows for the BTL flares. This is
similar to that  found for the on-disk flares for the same time frames, but
some on-disk flares are detected over longer time intervals (up to 20 hours) during
which the proton index increases gradually.

\end{enumerate}

We now present our  interpretation of these results with the goal of
constraining the emission and acceleration mechanisms and sites.

\subsection{SEP onset times}
Since the spectral information available for the Jan14
flare is  limited we cannot constrain the acceleration, transport or radiation
processes. However this flare is remarkable because the AR being located beyond
the western limb provided good magnetic connection  with the Earth and as a
result it was associated with a very strong SEP event and increased the
neutron flux  in the South Pole neutron monitor.  As presented in Figure~\ref{fig:SOL20140106OnsetTimes}, the acceleration time at the Sun for the SEP protons is 07:55 $\pm$ 0:05 UT which is in agreement with the SPR time of 07:47 $\pm$ 0:08 UT reported by~\citet{2014ApJ...790L..13T}. Unfortunately, the \Fermi-LAT was in the SAA at
the SPR time and exited at 07:55 UT. Therefore we can only provide an upper limit
on the time of acceleration of gamma-ray producing particles (presumably
protons)  at the Sun. However, \konus detected emission in the
20--78 keV band starting at 07:43 UT implying that the electrons were accelerated
at the Sun at 07:35 UT, roughly 12 minutes prior to the SPR.

Due to the location of the AR the other two flares were not magnetically
well connected to the
Earth. For the Sep14 flare the \goes proton fluxes begin to increase
roughly 9 hours after the flare start time. \stereo-B observed the flare in its entirety but provides proton fluxes only in three energy bands, which is not sufficient to
estimate the acceleration time at the Sun as we did for Jan14.
Thus, we do not have reliable SEP data to test whether the onset times
of the SEP and gamma-rays coincide, however we can compare the height of the CME
at the time of the gamma-ray onset with the CME height at SPR found for
Jan14. Based on the 2$^{nd}$ order fit of height as a function of time
(from the LASCO CME online catalog) we estimated that the CME was at $\sim$2.5\Rs at
the onset time of the \Fermi-LAT emission. We found a similar value for Oct13.
These heights are significantly smaller than the predicted values reported
in~\citet{2014ApJ...790L..13T} for the CME heights at SPR as a function of
source longitudes from  Solar Cycle-23. To test whether these CME heights are
compatible with the gamma-ray emission detected from the BTL
flares, it would require detailed simulations on particle transport in the solar
magnetic field. This study is beyond the scope of this paper and will be
addressed in a future work.

\subsection{Spatial Distributions}
\label{sec:position}

The \hsi images show that
the 20 to 50 keV  HXRs, due to  electron bremsstrahlung,  are emitted  from $\approx$30$\arcsec$ ($L \sim 2\times 10^9$ cm) and $\approx$50$\arcsec$  ($L\sim 3.5\times 10^9$ cm) sources from
top of loops extending to heights of $\sim 4\times 10^9$  and   $\sim 2\times 10^{10}$~cm above the photosphere for Oct13 and Sep14 flares, respectively.
The column
depth required for a particle of mass $m$ and kinetic energy $E$ (in
units of $mc^2$), or Lorentz factor $\gamma=E+1=(1-\beta^2)^{-1/2}$, to
lose all its  energy via Coulomb interactions is 
\begin{multline}
  \label{eq:CoulDepth}
    N_\coul=(4\pi r_0^2 \ln \Lambda)^{-1}(m/m_e)\int_0^E \beta^2 dE \\
    =5\times 10^{22}(m/m_e)E^2/\gamma\, {\rm cm}^{-2},
    \end{multline}
where $r_0$ is the classical electron radius and $\ln \Lambda\sim 20$ is the
Coulomb logarithm. From Equation~\ref{eq:CoulDepth} we find that the column depth required to stop the $> 50$ keV electrons that produce the HXRs detected by \hsi is $\sim 5\times 10^{20}$cm$^{-2}$. Assuming a density of $n<10^{10}$ cm$^{-3}$, the sizes of the loop-top sources imply a column depth of $N_{\rm LT}< 2$ and $3.5\times 10^{19}$ cm$^{-2}$ for the two HXR sources, respectively. This means that the energy loss time $\tau_\coul\sim 2 N_\coul/(vn)$ is more than 10 times longer than the crossing time $\tcross\equiv L/v=N_{\rm LT}/(vn)$. These values, together with the results reported in~\citet{2013ApJ...777...33C} suggest that the HXR emission from these two BTL flares are thin target sources.

If these loop-top sources were to be produced in a thick-target, which can come about in two  ways under extreme conditions not likely to be the case here. The first is in the strong diffusion limit, i.e., if the electrons are trapped because the scattering time, $\tau_{sc}$, is smaller than the crossing time, $\tau_{cross}$. Note that because $\tau_\coul> \tcross$, the scattering cannot be due to Coulomb interactions but it could be due to turbulence. The second possibility is in the weak diffusion limit $\tsc>\tcross$, where trapping can occur if the field lines converge strongly.  In this case we need $\tsc>\tau_\coul/\eta$
\footnote{According to~\citet{2001ApJ...549..402M} $\eta\sim \ln (B_0/B_L)$, the log of the ratio  of magnetic  field strengths at the middle and ends of the magnetic bottle.} (with proportionality constant $\eta$ increasing with
increasing field convergence~\citep{Vahe2016}).
These are somewhat extreme conditions requiring either a very high density (therefore increasing the scattering, presumably turbulence) or a somewhat strong field convergence and a ``clean" loop with low density and low level  of turbulence.  

{\it Thus, the HXR emission is most likely  due to electron bremsstrahlung
from a thin-target loop-top source.}

Based on the positions of the \Fermi-LAT $>$100~MeV emission centroids alone we cannot exclude that the gamma rays come from the loop-top source. In order to investigate this possibility we rely on Equation~\ref{eq:CoulDepth} to find the column depth required to stop $>350$ MeV protons (that produce $>100$ MeV photons) and find  $N_\coul(350 {\rm MeV})\sim 10^{25}$ cm$^{-2}$. This is much larger than the loop-top column depth $N_{\rm LT}\sim 2-3.5\times 10^{19}$ cm$^{-2}$ so that, in
the absence of trapping, we are again dealing with a thin target loop-top
source and a small energy loss  compared to what would be the case for
emission from a thick target photospheric source. Consequently, we would need much
larger energies of accelerated protons than those given in Table~\ref{tab:BTLQuantities} assuming a thick target scenario. 
The condition for proton trapping in the loop-top (or any coronal trap
region, for that matter) is also  more extreme  than that described above for
electrons and we would need escape times $10^5$ times longer than the crossing time. {\it Thus, the emission detected by the \Fermi-LAT  is most likely due to decay of pions produced by energetic protons interacting in a thick-target photospheric source.}

\subsection{Spectral Distributions}
\label{sec:spectra}

As shown in \citet{1971SoPh...17..412L} and \citet{1990ApJ...359..541M} in a thin target scenario the relation between the HXR spectral index, $\gamma$, and the number index of the energy spectrum of the emitting electrons\footnote{This should not be confused with electron flux, $F(E)=vN(E),$ index, which in non relativistic limit is $\d \ln F(E)/d\ln  E =\gamma-1.0$ used commonly. In the relativistic regime these indexes are equal  and as stated in Equation~\ref{eq:deltagamma} $\delta=\gamma$.} through bremsstrahlung processes, $\delta$, is
\begin{equation}
  \delta = \left\{
  \begin{array}{ll}
    \gamma - 0.5 & \text{non-relativistic case},\\
    \gamma   & \text{relativistic case}.
  \end{array}\right.
  \label{eq:deltagamma}
\end{equation}
For optically thin synchrotron radiation, the relation between $\delta$ and the microwave photon spectral index, $\alpha$, can be describe by the following relations for the relativistic \citep[e.g.][]{1979rpa..book.....R} and semi-relativistic \citep[e.g.][]{1985ARA&A..23..169D} cases

\begin{equation}
  \delta = \left\{
  \begin{array}{ll}
    1.1\alpha + 1.36 & \text{semi-relativistic case},\\
    2.0\alpha +1 & \text{relativistic case}.
  \end{array}\right.
  \label{eq:synchr}
\end{equation}

We will use these relations in the following for the interpretation of the detected emission from the two brightest behind-the-limb flares reported in this work.
The spectra of the two flares in the 30 keV
to 10 MeV energy range are somewhat different so we discuss them separately.

\emph{Oct13:} As already shown in \citet{2015ApJ...805L..15P},
both \hsi and \Fermi-GBM detect  HXR spectra for Oct13 flare
that are relatively steep (best-fit value of power-law photon index $\gamma = 3.2\pm0.1$). However,
as shown in Figure~\ref{fig:SEDSOL20131011} there is a tentative indication that the spectrum
may harden into a nearly flat part in the 1--10 MeV range  requiring either a
similar hardening of the spectrum of the accelerated electrons, as e.g.
described in~\citet{1997ApJ...489..358P}, or a contribution from  de-excitation nuclear lines. We found that the inclusion of a 2.223 MeV
broad neutron capture line and the nuclear de-excitation lines did not improve the fit significantly and that the best-fit model for the emission observed by \Fermi-GBM is a single power law.

As described in Section~\ref{sec:position} the thin target is most likely, in which case following Equation~\ref{eq:deltagamma} for the non-relativistic case ($<$100 keV) the index of the electron number density $\delta \sim 2.7$. 
The radio spectrum observed by RSTN, shown in Figure~\ref{fig:RadioSpecOct13}, indicates  microwave flux of 60
solar  flux units ($6\times 10^5$ Jy) that peaks at frequency $\nu\sim 5$ GHz
and declines as $\nu^{-\alpha}$ with the index $\alpha\sim 2$ above this
frequency. Based on the relations described in Equation~\ref{eq:synchr} we find that the required power-law index for the number density of electrons to be $\delta\sim 5$ or $\delta\sim 3.5$ in the relativistic or semi-relativistic regime, respectively.

These values are steeper than the index $\delta \sim 2.7$ required in the non-relativistic HXR range, indicating a steepening of the electron spectrum above  1 MeV. {\it We therefore conclude that the GBM flux in the energy range 300 keV - 10 MeV is probably not due to the accelerated electrons and no extra component is expected.}

\emph{Sep14:}
The best-fit model for the combined HXR \Fermi-GBM gamma-ray emission from the Sep14 flare is a single power law with an exponential decay with photon index 2.06$\pm$0.01 and a cut-off energy of 90$\pm$7 MeV. This is an extremely hard spectrum for a solar flare and if it is produced by a thin-target
electron-bremsstrahlung model as described above, it would require an electron number
 spectrum with index $\delta \sim 1.5$ in the non-relativistic range 
steepening to an index $\delta\sim 2$ in the relativistic range. However, as
noted above  inclusion of the 2.223 MeV line
improves the fit slightly, indicating a possible contribution from nuclear
de-excitation lines so the steepening of the electron spectrum could
be larger.

The radio spectrum of this flare, shown in Figure~\ref{fig:RadioSpecOct13} is also harder. The emission detected by the
San Vito station  has an index $\alpha\sim 1$ above 1 GHz. Again assuming
optically thin synchrotron emission would require semi-relativistic electrons with electron power-law  index $\delta\sim 2.5$  indicating a slightly larger
steepening at high energies. If the magnetic field is lower, which
could  be the case because of the larger loop, the steepening would be
even larger ($\delta\sim 3$). {\it Thus, for the Sep14 flare a power-law spectrum, steepening at higher energies, is required to explain the HXR and microwave observations from the loop-top region. This plus a pion decay component provides an acceptable fit to the Fermi data from 30 keV to several GeV. }

\subsection{Acceleration Site and Mechanism} 
\label{sec:accl}

The above interpretation of the HXR and microwave emissions indicates that the
electrons responsible for these emissions are most likely accelerated in the
reconnection region above the loop-top source either by turbulence~\citep{petrosian2012stochastic} or in merging islands manifested in the particle-in-cell simulations~\citep{2012APS..APR.K1022D}. The electrons are trapped  in the loop-top region long
enough (longer than the crossing time of a fraction of second) to produce
detectable bremsstrahlung and synchrotron radiation, but most of their energy
is lost in thick-target footpoints hidden from near Earth instruments because
of high optical depths~\citep{2015ApJ...805L..15P}. This scenario can also
explain the HXR emission from  a BTL flare reported by \citet{KruckerS_Coronal60keV_2007ApJ...669L..49K}. The
steepening of the electron spectrum in the semi-relativistic regime is due to a
decrease in the acceleration rate and not to transport or energy loss effects.

If the \Fermi-LAT emission is from a thick-target photospheric site, then this emission site is located on the visible part of the disk. It can not come from the occulted AR. The fact that accelerated protons reach the on-disk emission site provides strong evidence that the acceleration site is the CME environment, as suggested by \citet{cliv93} and \citet{2015ApJ...805L..15P}. They would diffuse across a wide range of magnetic fields, some connected to the AR behind the limb and some to the visible side of the Sun.  These ions will most likely
come from the down-stream region of the CME shock while SEPs escape
the upstream region. This can account for the differences in their spectral and
temporal characteristics. It is then possible that the discrepancy between
the spectral characteristics in 1--10 MeV  for the GBM and radio observations
can be explained by lower-energy 1--100 MeV protons or electrons precipitating to
the visible disk side from the down-stream region and producing the 1--10 MeV
nuclear de-excitation line or bremsstrahlung emission below the chromosphere.
These and other possibilities, for example, production of the \Fermi-LAT emission by
relativistic electrons either in the CME shock, in reconnection regions on
current sheets behind the CME or trapped in a large loop with strong
convergence, will be discussed in future papers.

\acknowledgements
The $Fermi$ LAT Collaboration acknowledges support from a number of agencies and
institutes for both development and the operation of the LAT as well as
scientific data analysis. These include NASA and DOE in the United States,
CEA/Irfu and IN2P3/CNRS in France, ASI and INFN in Italy, MEXT, KEK, and JAXA in
Japan, and the K.~A.~Wallenberg Foundation, the Swedish Research Council and the
National Space Board in Sweden. Additional support from INAF in Italy and CNES
in France for science analysis during the operations phase is also gratefully
acknowledged. W.L. was supported by NASA HGI grant NNX16AF78G and LWS grant NNX14AJ49G. V.P, W.L and F.R.d.C are supported by NASA grants NNX14AG03G, NNX13AF79G and NNX12AO70G. Val. P and L.K. thanks the RFBR grant 15-02-03717 We thank Nariaka Nitta and Meng Jin for friutful discussions on SEPs and magnetic connectivity.
%
\newcommand{\noop}[1]{}


\end{document}